\documentclass[aps,prd,preprintnumbers,groupedaddress,nofootinbib,amssymb,eqsecnum,notitlepage]{revtex4-1}
\usepackage{here}
\usepackage{graphicx}
\usepackage{amsmath,amsthm,amssymb}
\usepackage{bm}
\usepackage{color}

\usepackage{amsfonts}
\usepackage{dcolumn}
\allowdisplaybreaks[1]

\begin{document}
\newcommand{\newc}{\newcommand}

\newcommand{\rk}[1]{{\color{red} #1}}
\newcommand{\ben}{\begin{eqnarray}}
\newcommand{\een}{\end{eqnarray}}
\newc{\be}{\begin{equation}}
\newc{\ee}{\end{equation}}
\newc{\ba}{\begin{eqnarray}}
\newc{\ea}{\end{eqnarray}}
\newc{\bea}{\begin{eqnarray*}}
\newc{\eea}{\end{eqnarray*}}
\newc{\tp}{\dot{\phi}}
\newc{\ttp}{\ddot{\phi}}
\newc{\nrhon}{n_I\rho_{I,n_I}}
\newc{\nrhocn}{n_c\rho_{c,n_c}}
\newc{\drho}{\delta \rho_I}
\newc{\drhoc}{\delta \rho_c}
\newc{\dphi}{\delta\phi}
\newc{\D}{\partial}
\newc{\ie}{{\it i.e.} }
\newc{\eg}{{\it e.g.} }
\newc{\etc}{{\it etc.} }
\newc{\etal}{{\it et al.}}
\newcommand{\nn}{\nonumber}
\newc{\ra}{\rightarrow}
\newc{\lra}{\leftrightarrow}
\newc{\lsim}{\buildrel{<}\over{\sim}}
\newc{\gsim}{\buildrel{>}\over{\sim}}
\newc{\aP}{\alpha_{\rm P}}
\newc{\delj}{\delta j}
\newc{\rhon}{\rho_{m,n}}
\newc{\rhonn}{\rho_{m,nn}}
\newc{\delrho}{\delta \rho_m}
\newc{\pa}{\partial}
\newc{\E}{{\cal E}}
\newc{\rH}{{\rm H}}
\newc{\ty}{\tilde{y}}

\title{
Scaling solutions and weak gravity in dark energy \\ 
with energy and momentum couplings}

\author{Luca Amendola$^{1}$ and Shinji Tsujikawa$^{2}$}

\affiliation{$^1$Institute of Theoretical Physics, Heidelberg University, 
Philosophenweg 16, 69120 Heidelberg, Germany}

\affiliation{$^2$Department of Physics, Faculty of Science, 
Tokyo University of Science, 1-3, Kagurazaka,
Shinjuku-ku, Tokyo 162-8601, Japan}

\begin{abstract}
We argue that the $\Lambda$CDM tensions of the Hubble-Lema\^itre expansion rate $H_0$ and the clustering normalization $\sigma_8$ can be eased, at least in principle, by considering
an interaction between dark energy and dark matter in such a way to induce a small and positive early effective equation of state and a weaker gravity. 
For a dark energy scalar field $\phi$ interacting with dark matter through an exchange 
of both energy and momentum, we derive a general form of the Lagrangian 
allowing for 
the presence of scaling solutions. In a subclass of such interacting theories,
we show the existence of a scaling $\phi$-matter-dominated-era ($\phi$MDE)  which can potentially alleviate the $H_0$ tension by generating an effective high-redshift equation of state. 
We also study the evolution of perturbations for a model with $\phi$MDE followed 
by cosmic acceleration and find that the effective gravitational coupling relevant 
to the linear growth of large-scale structures can be smaller than the Newton 
gravitational constant $G$ at low redshifts.
The momentum exchange between dark energy and dark matter plays a crucial 
role for realizing weak gravity, while the energy transfer is also required 
for the existence of $\phi$MDE.

\end{abstract}

\date{\today}

\pacs{04.50.Kd, 95.36.+x, 98.80.-k}

\maketitle

\section{Introduction}
\label{introsec}

Testing gravitational interactions with cosmological and astrophysical observations 
is one of the most interesting area 
of current research. Several large-scale surveys that will contribute to this task 
are underway or planned for the next few years \cite{Abell:2009aa,Rawlings:2011dd,Laureijs:2011gra,Amendola:2012ys,Levi:2013gra,Riess:2018}. 
If there is a deviation from Einstein gravity, then new observable phenomena are expected 
to appear, including stronger or weaker gravitational clustering and lensing, scale-dependent perturbation growth, screening of fifth forces, breaking of the equivalence principle, 
anomalous propagation of gravitational waves, and so on. 

In order to predict these new effects, one often analyses the most general gravity 
theories with second-order equations of motion--like 
scalar-tensor \cite{Horndeski,Def,KYY,Char11,DKT11,DT12}, 
vector-tensor  \cite{Heisenberg,Tasinato1,DeFelice:2016yws,DeFelice:2016uil}, 
massive gravity \cite{Hassan:2011zd,deRham:2010kj}
or, alternatively, proceeds by building effective Lagrangians that include all possible viable operators \cite{EFT1,EFT2,EFT3,EFT4,EFT5}. These approaches have the great virtue of being systematic, but they rapidly lead to a proliferation of operators that make difficult to extract definite predictions. The gravitational-wave event GW170817 \cite{GW170817}, 
together with its electromagnetic counterpart \cite{Goldstein}, put additional constraints 
on the allowed Lagrangian \cite{GW1,GW2,GW3,GW4,GW5,GW6}, 
but there is still degeneracy among different 
dark energy models. 
This is partly attributed to the fact that, for the theories in which the speed of 
gravity is equivalent to that of light, the effective gravitational coupling for 
cold-dark-matter (CDM) density perturbations is usually larger than the Newton gravitational 
constant $G$ in both scalar-tensor and vector-tensor 
theories \cite{Amendola17,Kase:2018aps,Nakamura:2018oyy}.

In this paper, we proceed in a different way as compared to the theoretical approach mentioned above.
We begin by identifying which are the physical effects we are interested in, 
and then build a general model that is expected to generate them. The physical effects 
we consider are motivated by the two most interesting discrepancies between 
the $\Lambda$CDM model and  
current observations, namely the $H_0$ and $\sigma_8$ tensions. 
Here, $H_0=100\,h$ km\,s$^{-1}$\,Mpc$^{-1}$ is today's Hubble-Lema\^itre 
expansion rate and 
$\sigma_8$ is the amplitude of matter perturbations within the comoving $8h^{-1}$ scale.

As is well-known, the $H_0$ tension arises because the estimate of $H_0$  from 
Cosmic Microwave Background (CMB) temperature anisotropies \cite{Planck2015} 
differs by more than 3$\sigma$ from the one based on its local measurements \cite{Riess:2018}. 
While there remains the possibility that this tension is 
due to unknown or uncontrolled systematic effects, the problem 
has been confirmed and exacerbated in recent observations \cite{Verde:2019ivm,Riess:2019cxk,Freedman:2019jwv,Reid:2019tiq}.
Similarly, the $\sigma_8$ tension, which is weaker than the $H_0$ tension, is due to the difference between the clustering normalization obtained from CMB and the one from shear lensing analyses \cite{weak1,weak2}, where the former favours  values 
of $\sigma_8$ larger than the latter.
In both cases, the CMB estimate depends on 
assuming  $\Lambda$CDM   as 
a model of dark energy and dark matter.
It is therefore natural to ask whether a suitable modification of the dark sector 
can lead to a better agreement with the data.

One way to reconcile, at least potentially, $H_0$ CMB results \cite{Planck2015} with a higher local value \cite{Riess:2018}, is to modify the evolution of the Universe between matter-radiation equality and cosmic acceleration, so that the effective early equation of state 
is slightly positive. In this case, one can show that the CMB acoustic peaks move to 
smaller angular scales, and therefore  a higher $H_0$ is needed to bring the model 
back to agreement with observations (although of course a full likelihood analysis 
is needed to assess the valid parameter space). 
One way to achieve this without altering the physical properties of matter is to couple dark matter to dark energy in such a way that a matter-dominated epoch 
with a small fraction of dark energy
is present. Indeed, for a canonical scalar-field dark energy model interacting with dark matter 
through the energy transfer characterized by a coupling constant $Q$ \cite{Wetterich,Amendola99}, 
there exists a scaling
$\phi$-matter-dominated-era ($\phi$MDE) during which the field 
density parameter $\Omega_\phi$ is constant ($\Omega_\phi=2Q^2/3$) \cite{Amendola99}. 
During the $\phi$MDE the scalar-field kinetic energy dominates over its potential energy, yielding a field equation of state $w_\phi=1$. 
In this case, a small and positive effective equation of state arises, such that $w_{\rm eff}=w_\phi\Omega_\phi=\Omega_\phi$.
A comparison with cosmological datasets has put upper limits to  $Q$ \cite{Pettorino2013,Ade:2015rim}.

In the models of Refs.~\cite{Wetterich,Amendola99}, the dark energy scalar field 
$\phi$ interacts with CDM 
(but not baryons) via a conformal rescaling of the metric. 
This type of couplings arises in Brans-Dicke theories \cite{BD} 
after a conformal transformation to the Einstein 
frame \cite{Amendola:1999qq,Khoury,Tsujikawa:2008uc}.
In the language of Schutz-Sorkin action \cite{Sorkin,Brown} 
describing CDM as a perfect fluid, 
the interacting Lagrangian for the models of Refs.~\cite{Wetterich,Amendola99} 
is of the form 
$L_{\rm int}=(e^{Q\phi/M_{\rm pl}}-1)\rho_m$ \cite{Frusciante:2018tvu},
where $\rho_m$ is the CDM density and $M_{\rm pl}$ is the 
reduced Planck mass.
Reflecting the fact that this interaction corresponds to the energy transfer, 
the effective gravitational coupling $G_{\rm eff}$ associated with the linear 
evolution of CDM perturbations is larger than $G$, 
such that $G_{\rm eff}=(1+2Q^2)G$ \cite{Amendola:2003wa}.

One problem for the models with $G_{\rm eff}>G$ is that the attractive gravitational 
force induced by the coupling increases the cosmic growth rate, which makes 
the $\sigma_8$ tension worse. 
Ideally, one would need a coupling such that the $\phi$MDE is accompanied 
or followed by weaker growth of matter perturbations. 
An effective gravitational coupling smaller than $G$ can be achieved for 
a model in which the scalar field $\phi$ interacts with CDM through 
a momentum transfer \cite{Pourtsidou:2013nha,Boehmer:2015kta,Boehmer:2015sha,Skordis:2015yra,
Koivisto:2015qua,Pourtsidou:2016ico,Dutta:2017kch,Linton,Kase:2019veo,Chamings:2019kcl}.
This type of interaction is based on the field derivative coupling with 
the CDM four velocity $u^{\mu}$ and it can be quantified by the scalar 
combination $Z=u^{\mu} \nabla_{\mu} \phi$, where $\nabla_{\mu}$ is 
the covariant derivative operator. 
In previous works, it was shown that the interacting Lagrangian such as 
$L_{\rm int} \propto Z^2$ \cite{Pourtsidou:2013nha,Pourtsidou:2016ico}
or $L_{\rm int} \propto X^{1-n/2} Z^n$ \cite{Kase:2019mox}, where 
$X=-\nabla^{\mu} \phi \nabla_{\mu} \phi/2$ 
and $n>0$, can realize $G_{\rm eff}<G$ for the theories without 
energy transfer (i.e., $Q=0$ for the notation used above).

In the Schutz-Sorkin action approach, it is straightforward to accommodate 
both energy and momentum couplings between 
perfect-fluid dark matter and scalar-field dark energy. 
For the CDM density $\rho_m$ that depends on its number density $n$, 
the interacting action may be expressed in the form,
\be
{\cal S}_{\rm int}=\int {\rm d}^4 x \sqrt{-g} \left[-f_1(\phi,X,Z) \rho_m (n)
+f_2(\phi,X,Z) \right]\,,
\label{action0}
\ee
where $g$ is the determinant of metric tensor $g_{\mu \nu}$, and 
$f_1$, $f_2$ are functions of $\phi$, $X$, $Z$. 
The first and second terms in the square bracket of Eq.~(\ref{action0}) 
characterize the energy and momentum transfers, respectively. 
Unlike the other models in which phenomenological coupling 
terms are added to the background equations 
by hand \cite{Dalal:2001dt,Zimdahl:2001ar,Chimento:2003iea,Wang1,Wei:2006ut,Amendola:2006dg,Guo:2007zk,Valiviita:2008iv,Gavela:2009cy,Wands,Kumar:2016zpg,DiValentino:2017iww,An:2017crg,Yang:2018euj,Pan:2019gop,DiValentino:2019ffd,Yang:2019uog}, 
the evolution of cosmological perturbations is unambiguously fixed 
in our interacting theory 
with the explicit action (\ref{action0}). 

The preceding discussion makes it clear that we need both scaling $\phi$MDE and 
weak gravity to alleviate the $H_0$ and $\sigma_8$ tensions.
As it will be demonstrated in this work, these requirements can be satisfied 
when both energy and momentum exchanges occur between dark energy and CDM. 
For this purpose, we first derive a general interacting Lagrangian for the existence 
of scaling solutions. For simplicity, we assume Einstein gravity with  
the scalar interacting action (\ref{action0}) and the Schutz-Sorkin action for CDM, 
while we leave baryons and radiation uncoupled. 
In this case, there is no difference for the propagation of 
gravitational waves in comparison to general relativity 
and no screening mechanism is needed.
We note that the function $f_2$ in Eq.~(\ref{action0}) also accommodates the scaling 
Lagrangian derived for k-essence with the functional dependence 
$f_2(\phi,X)$ \cite{Piazza,Tsuji04,Amendola:2006qi}. 

After obtaining the general scaling Lagrangian, we consider a subclass of models 
with $\phi$MDE and show that it is possible to realize 
$G_{\rm eff}<G$ 
at low redshifts even in the presence of both energy and momentum transfers.
The hope is that such interacting models can really ease both the $H_0$ and $\sigma_8$ tensions. 
We leave however the task of a full likelihood analysis with 
the current observational data for a future work.

\section{Coupled dark energy with a Lagrangian formulation}
\label{eomsec}

We consider a dark energy scalar field $\phi$ coupled to 
a barotropic perfect fluid described by a Schutz-Sorkin 
action \cite{Sorkin,Brown}.
The interacting Lagrangian is generally given by the form 
$L(n, \phi, X, Z)$ \cite{Pourtsidou:2013nha}, 
where $L$ is a function of the fluid number density $n$, 
the scalar field $\phi$ and its kinetic 
energy $X=-g^{\mu \nu} \nabla_{\mu}\phi 
\nabla_{\nu} \phi/2$, and 
$Z=u^{\mu} \nabla_{\mu} \phi$, with $u^{\mu}$ being 
the fluid four velocity. 
We separate the interacting Lagrangian $L$ into the 
sums of energy transfer $-f_1(\phi, X, Z) \rho_m(n)$ 
and momentum transfer $f_2(\phi, X, Z)$, where 
$\rho_m$ is the fluid density that depends on $n$. 
For CDM, the density has the linear dependence 
$\rho_m \propto n$. 
For the gravity sector, we consider the Einstein-Hilbert 
action described by the Lagrangian $(M_{\rm pl}^2/2)R$, 
where $R$ is the Ricci scalar.

The total action is then given by 
\be
{\cal S} = \int {\rm d}^4 x \sqrt{-g}\,\frac{M_{\rm pl}^2}{2}R
-\int {\rm d}^{4}x \left[
\sqrt{-g}\,\rho_m(n)
+ J^{\mu} \partial_{\mu} \ell  \right]
+\int {\rm d}^4x \sqrt{-g}\,L(n,\phi,X,Z)\,,
\label{action}
\ee
where  
\be
L(n,\phi,X,Z)=-f_1(\phi,X,Z) \rho_m (n)
+f_2(\phi,X,Z)\,.
\label{Lin}
\ee
The second integral in Eq.~(\ref{action}) corresponds to 
the Schutz-Sorkin action, where $n$ is related to the vector field $J^{\mu}$, as 
\be
n=\sqrt{\frac{g^{\mu \nu} J_{\mu} J_{\nu}}{g}}\,.
\ee
Unlike Refs.~\cite{DeFelice:2016yws,DeFelice:2016uil}, 
we do not take vector perturbations into account 
in the Schutz-Sorkin action, as they are nondynamical in scalar-tensor theories.
The fluid four-velocity $u_{{\mu}}$ is defined by 
\be
u_{{\mu}} \equiv \frac{J_{{\mu}}}{n\sqrt{-g}}\,,
\label{udef}
\ee
which satisfies the relation $u^{\mu} u_{{\mu}}=-1$. 
The scalar quantity $Z$ is expressed as $Z=g^{\mu \nu} J_{\mu} \nabla_{\nu} \phi/(n\sqrt{-g})$, while 
the scalar variable $\ell$ is a Lagrange multiplier. 
Since we are not modifying the Einstein-Hilbert action 
from general relativity, the speed of gravitational waves 
is equivalent to that of light.
 
Varying the action (\ref{action}) with respect to $\ell$, 
it follows that 
\be
\partial_{\mu} J^{\mu}=0\,.
\label{Jmure}
\ee
The fluid pressure is given 
by \cite{Pourtsidou:2013nha,Boehmer:2015kta,Boehmer:2015sha,DeFelice:2016yws,DeFelice:2016uil} 
\be
P_m=n \rho_{m,n}-\rho_m\,.
\ee
Substituting $J^{\mu}=n\sqrt{-g}\,u^{\mu}$ 
into Eq.~(\ref{Jmure}) and exploiting the properties 
$\partial_{\mu} (\sqrt{-g} u^{\mu})=\sqrt{-g} 
\nabla_{\mu} u^{\mu}$ and
$(\rho_m+P_m)\partial_{\mu} n=n \partial_{\mu} \rho_m$, 
we obtain
\be
u^{\mu} \partial_{\mu} \rho_m+
\left( \rho_m+P_m \right) 
\nabla_{\mu} u^{\mu}=0\,.
\label{conser1}
\ee
Variation of the action (\ref{action}) with respect to $J^{\mu}$ leads to 
\be
\partial_{\mu} \ell= 
u_{{\mu}} \left( 1+f_1 \right) \rho_{m,n}
-\frac{f_{1,Z} \rho_m-f_{2,Z}}{n} \left( 
\nabla_{\mu} \phi+Z u_{\mu} \right)\,,
\label{lc}
\ee
where we used the relation $\partial n/\partial J^{\mu}
=J_{{\mu}}/(n g)$ and the notations 
$\rho_{m,n} \equiv \partial \rho_m/\partial n$ and 
$f_{i,Z} \equiv \partial f_{i}/\partial Z$ with $i=1,2$.

For the variation of action (\ref{action}) with respect to 
$g^{\mu \nu}$, we employ the following properties,
\ba
\delta \sqrt{-g} 
&=& -\frac{1}{2} \sqrt{-g}\,
g_{\mu \nu} \delta g^{\mu \nu}\,,\\
\delta n &=& \frac{n}{2} \left( g_{\mu \nu} 
-u_{\mu} u_{\nu} \right) \delta g^{\mu \nu}\,,\\
\delta X &=&-\frac{1}{2} \nabla_{\mu} \phi
\nabla_{\nu} \phi\,\delta g^{\mu \nu}\,,\\
\delta Z &=& \left( \frac{1}{2} Z u_{\mu} 
u_{\nu} +u_{\mu}\nabla_{\nu} \phi \right) 
\delta g^{\mu \nu}\,,\\
\delta (J^{\mu} \partial_{\mu} \ell)
&=& \delta (g^{\mu \nu} J_{\mu} \partial_{\nu} \ell)
=J_{\mu} \partial_{\nu} \ell\, \delta g^{\mu \nu}\,.
\ea
When the action is varied with respect to $g^{\mu \nu}$, we need to keep $J^\mu$ fixed 
to reproduce the standard matter energy-momentum tensor. 
Therefore, the vector $J^{\mu}$ should not be replaced with $n\sqrt{-g}\,u^\mu$ 
when varying the action.
Then, we obtain the following covariant equations of motion, 
\be
M_{\rm pl}^2 G_{\mu\nu}=\left( 1+f_1 \right) T^{(m)}_{\mu\nu}+T_{\mu\nu}^{(\phi)}\,,
\label{Ein}
\ee
where $G_{\mu \nu}$ is the Einstein tensor, and
\ba
T^{(m)}_{\mu\nu} 
&=& \left( \rho_m+P_m \right) 
u_{\mu} u_{\nu}+P_m g_{\mu \nu}\,,
\label{Tmmun}\\
T^{(\phi)}_{\mu \nu} 
&=& f_2 g_{\mu \nu}-\rho_m \left( f_{1,X} 
\nabla_{\mu} \phi \nabla_{\nu} \phi
+f_{1,Z}Z u_{\mu} u_{\nu} \right)
+f_{2,X}\nabla_{\mu} \phi \nabla_{\nu} \phi
+f_{2,Z} Z u_{\mu} u_{\nu}\,.
\ea
The matter energy-momentum tensor (\ref{Tmmun}) 
satisfies the continuity equation, 
\be
u^{\nu} \nabla^{\mu} T^{(m)}_{\mu\nu} 
=-\left[ u_{\mu} \partial^{\mu} \rho_m+
\left( \rho_m+P_m \right) 
\nabla^{\mu} u_{\mu} \right]=0\,,
\label{Tcon}
\ee
where we used Eq.~(\ref{conser1}) in the second equality. 
Taking the covariant derivative of Eq.~(\ref{Ein}),
we obtain
\be
\left( 1+f_1 \right) \nabla^{\mu} T_{\mu \nu}^{(m)}
+ T_{\mu \nu}^{(m)} \nabla^{\mu} f_1
+\nabla^{\mu} T_{\mu \nu}^{(\phi)}=0\,.
\label{Tcon2}
\ee
Multiplying Eq.~(\ref{Tcon2}) by $u^{\nu}$ and using the 
property (\ref{Tcon}), we find 
\be
u^{\nu} \nabla^{\mu}T^{(\phi)}_{\mu\nu}
=-u^{\nu} \nabla^{\mu} f_1\, T_{\mu \nu}^{(m)}\,.
\label{Tphicon}
\ee
We define the effective matter energy-momentum 
tensor, 
\be
\hat{T}^{(m)}_{\mu\nu} \equiv \left( 1+f_1 \right) 
T^{(m)}_{\mu\nu}\,,
\label{hatT}
\ee
which contains the effect of energy exchange between 
the scalar field and matter.
This quantity obeys
\be
u^{\nu}\nabla^{\mu} \hat{T}^{(m)}_{\mu\nu}
=+u^{\nu} \nabla^{\mu} f_1\, T_{\mu \nu}^{(m)}\,.
\label{Tmcon}
\ee
The signs on the right hand sides of Eqs.~(\ref{Tphicon}) and (\ref{Tmcon}) 
are opposite to each other, which shows the energy exchange 
between the scalar field and matter.
The coupling $f_2$, which corresponds to the momentum 
transfer, does not appear on the right hand sides of 
Eqs.~(\ref{Tphicon}) and (\ref{Tmcon}). 
However, the momentum transfer between the scalar field 
and matter occurs through Eq.~(\ref{Tcon2}).

Let us consider the flat 
Friedmann-Lema\^itre-Robertson-Walker (FLRW) 
background given by the line element 
${\rm d}s^2=-{\rm d}t^2+a^2(t) \delta_{ij} {\rm d}x^i {\rm d} x^j$, 
where $a(t)$ is the time-dependent scale factor.
On this background we have $u^{\mu}=(1,0,0,0)$ and 
$\nabla_{\mu} u^{\mu}=3H$, where $H=\dot{a}/a$ 
is the Hubble-Lema\^itre expansion rate and a dot represents 
a derivative with respect to the cosmic time $t$.
{}From Eq.~(\ref{conser1}), we obtain
\be
\dot{\rho}_m+3H \left( \rho_m+P_m \right)=0\,,
\label{con}
\ee
which corresponds to the conservation of total 
particle number ${\cal N} \equiv J^0=n a^3$. 
{}From the (00) and $(ii)$ components of Einstein equations (\ref{Ein}), 
it follows that 
\ba
& & 
3M_{\rm pl}^2 H^2=
\left( 1+f_1 \right) \rho_m+\rho_{\phi}\,,
\label{Eq00}\\
& & 
M_{\rm pl}^2 \left( 2\dot{H}+3H^2 
\right)=-\left( 1+f_1 \right) P_m-P_{\phi}\,,
\label{Eq11}
\ea
where 
\ba
\rho_{\phi} &\equiv& -\rho_m f_{1,X}\dot{\phi}^2 
-\rho_m f_{1,Z} \dot{\phi}-f_2+f_{2,X} \dot{\phi}^2
+f_{2,Z} \dot{\phi}\,,\\
P_{\phi} &\equiv& f_2\,.
\ea
{}From Eq.~(\ref{hatT}), we introduce the matter  
density and pressure containing the effect of energy transfer, as
\be
\hat{\rho}_m \equiv (1+f_1) \rho_m\,,\qquad 
\hat{P}_m \equiv  (1+f_1) P_m\,.
\ee
{}From Eqs.~(\ref{Tphicon}) and (\ref{Tmcon}), we have 
\ba
& & 
\dot{\rho}_{\phi}+3H \left( \rho_{\phi}
+P_{\phi} \right)=-\dot{f}_1 \rho_m\,,\label{Eq22} 
\\
& &
\dot{\hat{\rho}}_m+3H \left( \hat{\rho}_m
+\hat{P}_m \right)=+\dot{f}_1 \rho_m\,,
\label{con2}
\ea
whose right hand sides explicitly show the energy exchange between 
the scalar field and matter.

For  later convenience, we introduce the time-dependent 
density parameters,
\be
\Omega_m \equiv \frac{(1+f_1)\rho_m}{3M_{\rm pl}^2 H^2}\,,
\qquad
\Omega_\phi \equiv \frac{\rho_{\phi}}{3M_{\rm pl}^2 H^2}\,.
\label{Omega}
\ee
The form of matter is not  specified 
until the end of Sec.~\ref{fixedsec}, 
after which we take CDM, baryons, and 
radiation into account as perfect fluids.
From Eq.~(\ref{Eq00}), it follows that 
\be
\Omega_m+\Omega_{\phi}=1\,.
\ee
We also define the equations of state, 
\be
w_m \equiv \frac{\hat{P}_m}{\hat{\rho}_m}
=\frac{P_m}{\rho_m}\,,\qquad 
w_{\phi} \equiv \frac{P_{\phi}}{\rho_{\phi}}\,,
\ee
as well as the effective equation of state, 
\be
w_{\rm eff} \equiv w_m \Omega_m+
w_{\phi} \Omega_{\phi}= -1-\frac{2\dot{H}}{3H^2}\,.
\ee
The condition for cosmic acceleration to occur corresponds  
to $w_{\rm eff}<-1/3$.

\section{Scaling Lagrangian}
\label{scalingsec}

In this section, we derive the Lagrangian $L$ allowing for the existence 
of scaling solutions characterized by nonvanishing 
constants $\Omega_m$ and $\Omega_{\phi}$.
Notice that we are not requiring that the entire cosmic evolution 
has the scaling behavior, but just that scaling epochs exist.
Since  $(1+f_1) \rho_m \propto H^2$ and  $\rho_{\phi} \propto H^2$ in the scaling regime, 
it follows that 
\ba
\frac{\dot{f}_1}{H(1+f_1)} &=&-3 \left( 
w_{\rm eff}-w_m \right)\,,\label{scare1}\\
\frac{\dot{\rho}_{\phi}}{H \rho_{\phi}}
&=&-3(1+w_{\rm eff})\,,\label{scare2}
\ea
where we used Eq.~(\ref{con}). 
We will consider the case in which $w_{\rm eff}$, $w_m$, and 
$w_{\phi}$ are constant along the scaling solution. 
Then, all the terms in Eq.~(\ref{Eq11}) are proportional to $H^2$.
Since $\rho_{\phi} \propto P_{\phi} \propto H^2$, 
each term on the left hand side of Eq.~(\ref{Eq22}) is 
proportional to $H^3$. On using Eq.~(\ref{scare1}), 
there is also the dependence 
$\dot{f}_1 \rho_m \propto H (1+f_1) \rho_m \propto H^3$, 
so the scaling solution is consistent with Eq.~(\ref{Eq22}) 
as well.

If we consider a canonical scalar field given by the Lagrangian 
$f_2(\phi,X)=X-V(\phi)$, the field density $\rho_{\phi}$ 
contains the kinetic energy $\dot{\phi}^2/2$. 
In this case, the scaling solution satisfies 
the relation $\dot{\phi} \propto H$. 
To accommodate this model as a special case, 
we will derive the Lagrangian 
allowing for scaling solutions characterized by 
\be
\frac{\dot{\phi}}{HM_{\rm pl}}=\alpha\,,
\label{alpha}
\ee
where $\alpha$ is a dimensionless constant.
Then, the time derivative of $f_1(\phi,X,Z)$ yields
\be
\dot{f}_1=H \left[ \alpha M_{\rm pl} f_{1,\phi} 
-3(1+w_{\rm eff}) \left( X f_{1,X}+\frac{1}{2} 
Z f_{1,Z} \right) \right]\,.
\label{dotf1}
\ee
Substituting Eq.~(\ref{dotf1}) into Eq.~(\ref{scare1}), 
we obtain the partial differential equation, 
\be
Xf_{1,X}+\frac{1}{2}Zf_{1,Z}-\frac{1}{\lambda}M_{\rm pl} 
f_{1,\phi}+\frac{Q}{\lambda} \left( 1+f_1 \right)=0\,,
\label{pdif1}
\ee
where 
\be
\lambda \equiv \frac{3(1+w_{\rm eff})}{\alpha}\,,\qquad 
Q \equiv -\frac{3(w_{\rm eff}-w_m)}{\alpha}\,.
\ee
The integrated solution to Eq.~(\ref{pdif1}) is given by 
\be
f_1=e^{Q\phi/M_{\rm pl}} g_1 \left(Y_1,Y_2 \right)-1\,,
\label{sca}
\ee
where $g_1$ is an arbitrary function of 
\be
Y_1=X e^{\lambda \phi/M_{\rm pl}}\,,\qquad 
Y_2=Z e^{\lambda \phi/(2M_{\rm pl})}\,.
\ee
Along the scaling solution, $\phi$, $a$, and $H$ evolve, respectively, as 
\be
\phi=\phi_0+\alpha M_{\rm pl} \ln a\,,\qquad 
a \propto \left( t-t_0 \right)^{2/[3(1+w_{\rm eff})]}\,,
\qquad 
H=\frac{2}{3(1+w_{\rm eff})(t-t_0)}\,,
\ee
where $\phi_0$ and $t_0$ are constants.
This shows that both $Y_1$ and $Y_2$ remain constant 
in the scaling regime.
One can also show that both
$\rho_m f_{1,X} \dot{\phi}^2$ and $ \rho_m f_{1,Z} \dot{\phi}$ 
in $\rho_{\phi}$ are proportional to $H^2$, so all the terms 
in Eq.~(\ref{Eq00}) associated with the coupling $f_1$ obey 
the scaling relation. In other words, each term in $\rho_{\phi}$ 
arising from $f_1$ satisfies the same relation as 
Eq.~(\ref{scare2}).

The coupling $f_2$, which is equivalent to $P_{\phi}$ 
and is also present as one of the terms in 
$\rho_{\phi}$, should satisfy the scaling property $f_2 \propto H^2$. 
This translates to the relation $\dot{f}_2/(Hf_2)=-3(1+w_{\rm eff})$ 
and hence $f_2$ obeys the partial differential equation,
\be
Xf_{2,X}+\frac{1}{2}Zf_{2,Z}-\frac{1}{\lambda}M_{\rm pl}f_{2,\phi}-f_2=0\,.
\label{pdif2}
\ee
This is integrated to give
\be
f_2=X g_2 (Y_1, Y_2)\,,
\label{sca2}
\ee
where $g_2$ is an arbitrary function of $Y_1$ and $Y_2$. 
On using the property that $Y_1$ and $Y_2$ are constant 
along the scaling solution, both $f_{2,X} \dot{\phi}^2$ 
and $f_{2,Z} \dot{\phi}$ in $\rho_{\phi}$ are proportional 
to $H^2$.
 
In summary, the Lagrangian allowing for the existence 
of scaling solutions is given by 
\be
L=-\left[ e^{Q\phi/M_{\rm pl}}
g_1 \left(Y_1,Y_2 \right)-1 \right] \rho_m (n)
+Xg_2 (Y_1, Y_2)\,,
\label{slag}
\ee
which contains arbitrary functions $g_1$ and $g_2$ 
with respect to $Y_1$ and $Y_2$. 
For the choice $g_1(Y_1, Y_2)=1$, the coupling $f_1$
corresponds to the one studied in Refs.~\cite{Frusciante:2018tvu,Kase:2019veo}, 
i.e., $f_1=e^{Q\phi/M_{\rm pl}}-1$, where the 
constant $Q$ characterizes the strength of energy exchange.
For k-essence theories with $f_2=f_2(\phi,X)$, the scaling Lagrangian 
reduces to $f_2=X g_2(Y_1)$, which agrees with that derived
in Refs.~\cite{Piazza,Tsuji04,Amendola:2006qi}.
The $Z$ dependence in $f_1$ and $f_2$ gives rise to the additional $Y_2$ 
dependence in the functions $g_1$ and $g_2$.

\section{Fixed points}
\label{fixedsec}

We derive the fixed points for interacting theories given 
by the action (\ref{action}) with the scaling Lagrangian (\ref{slag}). 
In doing so, we introduce the following dimensionless variables, 
\be
x \equiv \frac{\dot{\phi}}{\sqrt{6}HM_{\rm pl}}\,,\qquad 
y \equiv \frac{M_{\rm pl}e^{-\lambda \phi/(2M_{\rm pl})}}{\sqrt{3}H}\,.
\label{xydef}
\ee
Then, the background values of $Y_1$ and $Y_2$ 
are expressed, respectively, as 
\be
Y_1=\frac{x^2}{y^2}M_{\rm pl}^4\,,\qquad 
Y_2=\frac{\sqrt{2} x}{y}M_{\rm pl}^2\,,
\label{Y12}
\ee
and hence $Y_1=Y_2^2/2$. 
It is also convenient to define
\be
\epsilon_{\phi} \equiv \frac{\ddot{\phi}}{H\dot{\phi}}\,,\qquad 
\xi \equiv \frac{\dot{H}}{H^2}\,.
\label{epxi}
\ee
Then, the variables $x$ and $y$ obey
the differential equations,
\ba
x' &=& x \left( \epsilon_{\phi}- \xi \right)\,,\label{dxeq}\\
y' &=& -y \left( \frac{\sqrt{6}}{2}\lambda x 
+ \xi  \right)\,,\label{dyeq}
\ea
where a prime denotes a derivative with respect to 
$N=\ln a$. 
On using Eqs.~(\ref{Eq00})-(\ref{Eq22}), it follows that 
\ba
\hspace{-0.5cm}
\epsilon_{\phi} &=&
\frac{f_{1,\phi}\rho_m-f_{2,\phi}
+3H (f_{1,X} \dot{\phi}+f_{1,Z})P_m
+3H (f_{2,X} \dot{\phi}+f_{2,Z})
-(f_{1,X\phi} \rho_m-f_{2,X\phi})\dot{\phi}^2
-(f_{1,Z\phi} \rho_m-f_{2,Z\phi})\dot{\phi}}
{[f_{1,X}\rho_m-f_{2,X}+(f_{1,XX}\rho_m-f_{2,XX})\dot{\phi}^2
+2(f_{1,XZ} \rho_m-f_{2,XZ}) \dot{\phi}
+f_{1,ZZ}\rho_m-f_{2,ZZ}]H \dot{\phi}}\,,\\
\hspace{-0.5cm}
\xi &=& 
\frac{(f_{1,X}\dot{\phi}+f_{1,Z})\dot{\phi} \rho_m
-(f_{2,X} \dot{\phi}+f_{2,Z})\dot{\phi}-(1+f_1)(\rho_m+P_m)}
{2H^2M_{\rm pl}^2}\,.
\ea
For the theories given by the functions (\ref{sca}) and (\ref{sca2}), 
the quantities $\Omega_{\phi}$, 
$w_{\phi}$, and $w_{\rm eff}$ are expressed as 
\ba
\Omega_{\phi} &=&\frac{g_1 x^2 (g_2+2Y_1 g_{2,Y_1}
+Y_2 g_{2,Y_2})-2Y_1 g_{1,Y_1}-Y_2 g_{1,Y_2}}
{g_1-2Y_1 g_{1,Y_1}-Y_2 g_{1,Y_2}}=1-\Omega_{m}\,,\label{Ophi}\\
w_{\phi} &=& \frac{g_2 x^2}{\Omega_{\phi}}\,,\label{wp}\\
w_{\rm eff} &=& g_2 x^2+w_{m}\Omega_{m} \,.\label{we}
\ea
In Appendix, we will present the autonomous equations 
for the scaling Lagrangian (\ref{slag}) 
with $f_1$ and $f_2$ given by 
Eqs.~(\ref{sca}) and (\ref{sca2}), respectively.

The fixed points of the above dynamical system are characterized 
by constant values of $x$ and $y$,  which we denote 
$x_c$ and $y_c$, respectively. 
The scaling fixed point corresponds to $x_c \neq 0$ and 
$y_c \neq 0$, so it should satisfy 
\ba
& &
\epsilon_{\phi}=\xi\,,\label{fix1}\\
&&
\xi=-\frac{\sqrt{6}}{2} \lambda x_c\,.
\label{fix2}
\ea
{}From Eq.~(\ref{fix2}), we have 
\be
g_{2,Y_2}=\frac{\sqrt{6}\lambda  g_1 x_c-3(1-\Omega_{\phi})
[(1+w_m)g_1-2Y_1 g_{1,Y_1}-Y_2 g_{1,Y_2}]
-6 g_1 x_c^2 (g_2+Y_1 g_{2,Y_1})}{3g_1 x_c^2 Y_2}\,.
\label{fix3}
\ee
Substituting this relation into Eq.~(\ref{Ophi}) and solving it 
for $\Omega_{\phi}$, we obtain
\be
\Omega_{\phi}=1+\frac{3(1+g_2 x_c^2)
-\sqrt{6} \lambda x_c}{3w_m}\,.
\label{Ops}
\ee
By using Eqs.~(\ref{fix1}) and (\ref{fix3}) with Eq.~(\ref{Ophi}), 
it follows that 
\be
\left( g_1-2Y_1 g_{1,Y_1}-Y_2 g_{1,Y_2} \right)
\left[ 2(Q+\lambda) x_c-\sqrt{6} (1+w_m) \right] 
\left[ 3(1+g_2 x_c^2)-\sqrt{6}\lambda x_c \right]=0\,.
\label{fac}
\ee
Provided that $g_1$ is different from the specific form 
$g_1=\sqrt{Y_1}\,F (Y_2/\sqrt{Y_1})$, 
where $F$ is a function of $Y_2/\sqrt{Y_1}$, 
we have $g_1-2Y_1 g_{1,Y_1}-Y_2 g_{1,Y_2} \neq 0$. 
Then, there are two fixed points satisfying Eq.~(\ref{fac}): 
(a) scaling solution, and (b) scalar-field dominated point. 
In the following, we discuss the properties of 
them in turn.

\subsection{Scaling solution (a)}

The scaling fixed point corresponds to
\be
x_c=\frac{\sqrt{6} (1+w_m)}{2(Q+\lambda)}\,.
\label{xcs}
\ee
Substituting Eq.~(\ref{xcs}) into Eqs.~(\ref{Ops}), 
(\ref{wp}), and (\ref{we}), we obtain
\ba
\Omega_{\phi} 
&=& \frac{[2Q (Q+\lambda)+3(1+w_m)g_2](1+w_m)}
{2w_m (Q+\lambda)^2}\,,\label{Omephia} \\
w_{\phi}
&=& \frac{3w_m (1+w_m) g_2}
{3(1+w_m)g_2+2Q(Q+\lambda)}\,,\\
w_{\rm eff} &=&\frac{w_m \lambda-Q}{Q+\lambda}\,.
\label{weffge}
\ea
Due to the constancy of $g_2(Y_1,Y_2)$ along the scaling solution, the quantities 
(\ref{Omephia})-(\ref{weffge}) do not vary in time. 
In the limit that $Q \to 0$, both $w_{\phi}$ and $w_{\rm eff}$ 
reduce to $w_m$. In this case, the density of scalar field 
scales in the same manner as that of matter. 
The existence of nonvanishing coupling $Q$ leads  
$w_{\phi}$ and $w_{\rm eff}$ being different from $w_m$.

Instead of using $g_2$ in Eq.~(\ref{Omephia}), it is possible
to express $\Omega_{\phi}$ in terms of 
$f_{2,X}$, $g_{2,Y_2}$ 
as well as $g_1$ and its derivatives with respect to 
$Y_1$ and $Y_2$. 
In doing so, we substitute Eq.~(\ref{Ops}) and 
the relation $g_{2,Y_1}=(f_{2,X}-g_2)/Y_1$ into Eq.~(\ref{fix3}) 
and solve it for $g_2$. 
Eliminating the term $g_2$ from Eq.~(\ref{Omephia}), we obtain
\be
\Omega_{\phi}=\frac{[2Q(Q+\lambda)+3(1+w_m)(2f_{2,X}+Y_2 g_{2,Y_2})]
(1+w_m)g_1-2(Q+\lambda)^2 (2Y_1 g_{1,Y_1}+Y_2 g_{1,Y_2})}
{2(Q+\lambda)^2[(1+w_m)g_1-2Y_1 g_{1,Y_1}-Y_2 g_{1,Y_2}]}\,.
\label{Opge}
\ee
If $g_1$ depends on neither $Y_1$ 
nor $Y_2$, we have
$\Omega_{\phi}=[2Q(Q+\lambda)+3(1+w_m)(2f_{2,X}+Y_2 
g_{2,Y_2})]/[2(Q+\lambda)^2]$. 
In k-essence where $g_2$ depends on $Y_1$ alone, 
the field density parameter further 
reduces to $\Omega_{\phi}=[Q(Q+\lambda)+3(1+w_m)
f_{2,X}]/(Q+\lambda)^2$. 
This coincides with the result
derived in Refs.~\cite{Tsujikawa:2006mw,Amendola:2006qi}.
In quintessence ($f_2=X-V(\phi)$) with the exponential 
potential $V(\phi)=V_0 e^{-\lambda \phi/M_{\rm pl}}$, i.e., 
the choice $g_2=1-V_0/Y_1$, we reproduce the value 
$\Omega_{\phi}=3(1+w_m)/\lambda^2$ for $Q=0$ \cite{CLW}.
In this case, the scaling radiation and matter eras 
in which $\Omega_{\phi}$ is subdominant to $\Omega_m$ can be 
realized for $|\lambda| \gg 1$ \cite{Barreiro,CST06}.

The field density parameter (\ref{Opge}) contains the dependence 
of both $Y_1$ and $Y_2$ in the functions $g_1$ and $g_2$, so 
it is the generalization of scaling solutions already known 
in literature. However, the effective equation of state (\ref{weffge}) 
is not subject to modifications compared to that derived 
in Refs.~\cite{Amendola99,Tsujikawa:2006mw,Amendola:2006qi}.
For $|\lambda| \gg 1$ and $|Q|<{\cal O}(1)$, $w_{\rm eff}$ is close to 
$w_m$. In this case, the scaling solution with $\Omega_{\phi} \ll 1$ 
can be used during the radiation or matter eras. 
For $|Q|$ larger than the order of $|\lambda|$, it is possible 
for the scaling solution to satisfy the condition of cosmic 
acceleration ($w_{\rm eff}<-1/3$). 
However, the coupling $|Q|$ is typically constrained to be 
smaller than  0.1 for the consistency with CMB 
measurements \cite{Ade:2015rim}, 
in which case the realization of scaling accelerated attractor 
with $\Omega_{\phi} \simeq 0.7$ is difficult \cite{Amendola:2006qi}.
For $|Q|<{\cal O}(0.1)$ and $|\lambda|<{\cal O}(1)$, 
the fixed point relevant to late-time cosmic acceleration 
is  the scalar-field dominated point 
discussed later in Sec.~\ref{pointBsec}.

\subsection{Scalar-field dominated point (b)}
\label{pointBsec}

{}From Eq.~(\ref{fac}), there exists the other fixed point satisfying 
\be
g_2=\frac{\sqrt{6} \lambda x_c-3}{3x_c^2}\,.
\label{g2re1}
\ee
Then, from Eqs.~(\ref{wp}), (\ref{we}), and (\ref{Ops}), 
we have 
\ba
& &
\Omega_{\phi}=1\,,\\
& &
w_{\rm eff}=w_{\phi}=-1+\frac{\sqrt{6} 
\lambda x_c}{3}\,.
\label{wep}
\ea
This scalar-field dominated point can be responsible for the late-time 
cosmic acceleration under the condition 
\be
\lambda x_c < \frac{\sqrt{6}}{3}\,.
\ee
{}From Eq.~(\ref{fix3}), there is also the following relation 
\be
g_2+Y_1 g_{2,Y_1}+\frac{1}{2} Y_2 g_{2,Y_2} 
=\frac{\lambda}{\sqrt{6}x_c}\,.
\label{g2re2}
\ee
For a given function $g_2(Y_1, Y_2)$, the values of $x_c$ and $y_c$ at the 
scalar-field dominated point are obtained 
by solving Eqs.~(\ref{g2re1}) and (\ref{g2re2}) together with 
the relations given in Eq.~(\ref{Y12}). 
On using Eq.~(\ref{g2re2}), we can express Eq.~(\ref{wep}) in the form 
\be
w_{\rm eff}=w_{\phi}=-1+\frac{2\lambda^2}
{3(2f_{2,X}+Y_2 g_{2,Y_2})}\,,
\label{weffb}
\ee
where $f_{2,X}=g_2+Y_1 g_{2,Y_1}$. 
In k-essence without the $Y_2$ dependence in $g_2$, 
we recover the value 
$w_{\rm eff}=w_{\phi}=-1+\lambda^2/(3f_{2,X})$ 
derived in Refs.~\cite{Tsujikawa:2006mw,Amendola:2006qi}. 
From Eq.~(\ref{weffb}), we find that, for $\lambda$ closer 
to 0, $w_{\rm eff}$ and $w_{\phi}$ approach $-1$.

\subsection{Kinetic fixed points}
\label{kinesec}

Let us derive other fixed points for the dynamical system 
given by Eqs.~(\ref{dxeq})-(\ref{dyeq}). 
The $\phi$MDE corresponds to the scaling solution with kinetic 
domination satisfying 
\be
y_c=0\,.
\ee
Since the quantities $Y_1$ and $Y_2$ diverge at $y_c=0$,
the functions $g_1$ and $g_2$ should take the following forms 
to avoid the divergence of background equations, 
\ba
g_1 (Y_1, Y_2) &=& 
b_0+\sum_{i>0} \left( b_i Y_1^{-i}+ 
\tilde{b}_i Y_2^{-i} \right)
+2^{1-m/2} \mu \frac{Y_2^m}{Y_1^{m/2}}
+\sum_{i>0,j<2i} 
\mu_i \frac{Y_2^j}{Y_1^i} \,, \label{g1e} \\
g_2 (Y_1, Y_2) &=& 
c_0+\sum_{i>0} \left( c_i Y_1^{-i}+ 
\tilde{c}_i Y_2^{-i} \right)
+2^{1-m/2} \beta \frac{Y_2^m}{Y_1^{m/2}}
+\sum_{i>0,j<2i} 
\beta_i \frac{Y_2^j}{Y_1^i} \,, \label{g2e} 
\ea
where $b_0$, $b_i$, $\tilde{b}_i$, $\mu$, $\mu_i$,  $c_0$, $c_i$, $\tilde{c}_i$, $\beta$, $\beta_i$,
and $m$ are constants.
Due to the relation $Y_1=Y_2^2/2$ at the background level, 
the third terms on the right hand sides of Eqs.~(\ref{g1e}) and (\ref{g2e}) are constant.
The last terms in Eqs.~(\ref{g1e}) and (\ref{g2e}) do not 
diverge for $i>0$ and $j<2i$.

In the following, we will set $c_0=1$ without loss of 
generality.
We substitute Eqs.~(\ref{g1e})-(\ref{g2e}) and their $Y_1$, $Y_2$ 
derivatives into Eq.~(\ref{Ophi})-(\ref{we}), use the relations 
(\ref{Y12}), and finally take the limit $y \to 0$. 
This process leads to 
\ba
& &
\Omega_{\phi}= q_s x^2 =1-\Omega_m\,,
\label{Omeki}\\
& &
w_{\phi}=1\,,\\
& &
w_{\rm eff}
=w_m-(w_m-1) q_s x^2 \,,
\label{wki}
\ea
where 
\be
q_s \equiv 1+2\beta\,.
\ee
The autonomous Eq.~(\ref{dxeq}) reduces to 
\be
x'=-\frac{1}{2q_s} \left[ 3q_s (w_m-1)x-\sqrt{6}Q 
\right] \left( q_s x^2-1 \right)\,,
\label{dxeq2}
\ee
while Eq.~(\ref{dyeq}) is automatically satisfied. 
We note that the constants appearing in the expression of 
$g_1 (Y_1, Y_2)$ in Eq.~(\ref{g1e}) do not affect 
Eq.~(\ref{dxeq2}).
It is also interesting to observe that, for $w_m=0$,  there is the relation $w_{\rm eff}=\Omega_{\phi}=q_s x^2$. 
Indeed, this relation holds in all the kinetic scaling 
solutions identified so far, see, for instance, Refs.~\cite{Amendola:2006qi,Amendola:2018ltt,Frusciante:2018tvu,Frusciante:2018aew}. 
It is this phase that induces a small and positive effective equation of state that can help alleviating the $H_0$ tension. On the other hand, this implies that a kinetic scaling solution associated with the matter dominance will be strongly constrained by CMB observations of the distance to last scattering. 
Leaving aside the $H_0$ tension, one possibility to implement $w_{\rm eff}=0$, and therefore to ease CMB distance constraints, would be to introduce a non-vanishing $w_m=\Omega_\phi/(\Omega_\phi-1)$. 
This possibility is not pursued here 
but left for future work. 
 
{}From Eq.~(\ref{dxeq2}), there are the following fixed points.

\subsubsection{$\phi$MDE {\rm (c)}}

The $\phi$MDE corresponds to one of the solutions to 
Eq.~(\ref{dxeq2}), i.e., 
\be
x_c=\frac{\sqrt{6}Q}{3 q_s (w_m-1)}\,.
\label{xcc}
\ee
{}From Eqs.~(\ref{Omeki}) and (\ref{wki}), we obtain
\ba
& &
\Omega_{\phi}=\frac{2Q^2}{3 q_s (w_m-1)^2}
=1-\Omega_m\,,
\label{Omekic}\\
& &
w_{\rm eff}
=w_m-\frac{2Q^2}{3 q_s (w_m-1)}\,.
\label{wkic}
\ea
This means that the $\phi$MDE is a scaling solution 
besides the fixed point (a). 
For nonrelativistic matter ($w_m=0$), it follows that 
$\Omega_{\phi}=w_{\rm eff}=2Q^2/(3q_s)$. 
The standard matter-dominated epoch 
with $\Omega_{\phi}=w_{\rm eff}=0$ 
is modified by the nonvanishing coupling $Q$. 
In comparison to theories with the coupling $Q$ alone, the coupling 
$\beta$ gives the additional contribution to 
 $x_c$, $\Omega_{\phi}$, and $w_{\rm eff}$.

\subsubsection{Purely kinetic solutions {\rm (d1), (d2)}}

The other solution to Eq.~(\ref{dxeq2}) corresponds to 
purely kinetic points (d1), (d2) satisfying
\be
x_c=\pm \frac{1}{\sqrt{q_s}}\,,
\ee
whose existence  requires that $q_s>0$. 
{}From Eqs.~(\ref{Omeki}) and (\ref{wki}), we have 
\be
\Omega_{\phi}=1\,,\qquad w_{\rm eff}=1\,.
\ee
The points (d1), (d2) are not relevant to 
radiation/matter eras or the epoch of cosmic acceleration.

\vspace{0.3cm}
In summary, the fixed points relevant to the cosmological 
evolution after radiation-matter equality are the $\phi$MDE (c) and the scalar-field dominated point (b). 
The coupling $Q$ associated with the energy transfer is crucially required for 
the existence of $\phi$MDE. 
The coupling $\beta$ associated with the momentum transfer also affects 
the values of  $\Omega_{\phi}$ and $w_{\rm eff}$ on the $\phi$MDE.

\section{Background cosmology for a model with $\phi$MDE}
\label{backsec}

In this section, we study the background cosmological dynamics 
for the model with 
\ba
& &
g_1(Y_1,Y_2)=1\,,
\label{eq:simpg1}\\
& &
g_2(Y_1,Y_2)=1-\frac{V_0}{Y_1}
+2^{1-m/2} \beta \frac{Y_2^m}{Y_1^{m/2}}\,,
\label{eq:simpg2}
\ea
where $V_0$ is a positive constant. 
Although this is a simple  model, it contains nevertheless 
all the new phenomenology we wish to consider. 
We consider the coupling only between the scalar field and CDM, 
where the CDM density $\rho_c$ depends on its number 
density $n_c$.
The interacting model given by Eqs.~(\ref{eq:simpg1})-(\ref{eq:simpg2}) 
belongs to a subclass of the functions (\ref{g1e}) and (\ref{g2e}), 
so the $\phi$MDE is present besides the other 
fixed points derived in Sec.~\ref{fixedsec}.
In this case, the Lagrangian (\ref{Lin}) is given by
\be
L=-\left( e^{Q\phi/M_{\rm pl}}-1 \right) \rho_c (n_c)
+X-V_0 e^{-\lambda \phi/M_{\rm pl}}
+\beta \left( 2X \right)^{1-m/2} Z^m\,.
\label{Lagcon}
\ee
The canonical scalar field $\phi$ with the potential 
$V(\phi)=V_0 e^{-\lambda \phi/M_{\rm pl}}$ interacts with CDM 
through the energy-transfer 
$-\left( e^{Q\phi/M_{\rm pl}}-1 \right) \rho_c (n_c)$ 
and the momentum-transfer  
$\beta \left( 2X \right)^{1-m/2} Z^m$. 
Without the coupling $Q$, the interactions of the forms 
$Z^2$ \cite{Pourtsidou:2013nha,Pourtsidou:2016ico,Chamings:2019kcl} 
or $Z^n$ \cite{Linton,Kase:2019mox} were already studied 
in the literature. 
In the following, we are going to investigate the cosmological 
dynamics in the presence of the two nonvanishing coupling 
constants $Q$ and $\beta$.

Besides CDM with the vanishing pressure ($P_c=0$), 
we also take baryons (energy density $\rho_b$ with 
vanishing pressure) and radiation (energy density 
$\rho_r$ and pressure $P_r=\rho_r/3$) into account 
to study the dynamics of background cosmology from 
the radiation era.
Neither baryons nor radiation are assumed to be 
coupled to the scalar field. 

Defining the variable $x$ as in Eq.~(\ref{xydef}), the dimensionless 
scalar field $\tilde{\phi} \equiv \phi/M_{\rm pl}$ obeys 
the differential equation,
\be
\tilde{\phi}'=\sqrt{6} x\,.
\ee
Instead of the variable $y$ defined in Eq.~(\ref{xydef}), 
we will use
\be
\tilde{y} \equiv \frac{\sqrt{V_0}}{M_{\rm pl}^2}y
=\sqrt{\frac{V_0}{3}} \frac{e^{-\lambda \phi/(2M_{\rm pl})}}
{M_{\rm pl} H}\,.
\label{tildey}
\ee
The density parameters of scalar field 
and matter components are given by 
\be
\Omega_{\phi}=q_s x^2+\tilde{y}^2\,,
\qquad
\Omega_c=\frac{e^{Q\phi/M_{\rm pl}} \rho_c}{3M_{\rm pl}^2 H^2}\,,
\qquad 
\Omega_b=\frac{\rho_b}{3M_{\rm pl}^2 H^2}\,,
\qquad 
\Omega_r=\frac{\rho_r}{3M_{\rm pl}^2 H^2}\,.
\ee
As we will see in Sec.~\ref{persec}, the no-ghost condition 
of scalar-field perturbation requires that 
\be
q_s=1+2\beta>0\,.
\label{noghost}
\ee
{}From Eq.~(\ref{Eq00}), (\ref{Eq11}) and (\ref{Eq22}), we have
\ba 
\Omega_c &=& 
1-\Omega_{\phi}-\Omega_b-\Omega_r\,,\\
\xi  &=& 
-3q_s x^2-\frac{3}{2}\Omega_c-\frac{3}{2}\Omega_b
-2\Omega_r\,,\\
\epsilon_{\phi}
&=& -3+\frac{\sqrt{6}}{2q_s x} 
\left( \lambda \tilde{y}^2-Q \Omega_{c} \right)\,,
\ea
where we recall that $\xi$ and $\epsilon_{\phi}$ are 
defined in Eq.~(\ref{epxi}).
We can reduce the background equations to the following 
autonomous system, 
\ba
x' &=& \frac{1}{2} x \left( 6 q_s x^2-6 +3\Omega_c
+3\Omega_b+4\Omega_r \right)
+\frac{\sqrt{6}}{2 q_s} \left( \lambda \tilde{y}^2
-Q \Omega_c \right)\,,\label{auto1}\\
\tilde{y}' &=& \frac{1}{2} \tilde{y} \left( 6 q_s x^2 
-\sqrt{6} \lambda x+3\Omega_c
+3\Omega_b+4\Omega_r \right)\,,\\
\Omega_b' &=& \Omega_b \left( 6 q_s x^2-3 
+3\Omega_c+3\Omega_b+4\Omega_r \right)\,,\\
\Omega_r' &=& \Omega_r \left( 6 q_s x^2-4 
+3\Omega_c+3\Omega_b+4\Omega_r \right)\,.
\label{auto4}
\ea
Notice that the background equations are independent of the power $m$. 
The scalar-field and effective equations of state are 
given, respectively, by 
\ba
w_{\phi} &=& \frac{ q_s x^2-\tilde{y}^2}
{q_s x^2+\tilde{y}^2}\,,\\
w_{\rm eff} &=& -1+2q_s x^2
+\Omega_c+\Omega_b+\frac{4}{3}\Omega_r\,.
\ea

Besides the scaling fixed point (a) derived in Sec.~\ref{fixedsec}, 
there exist the following three fixed 
points (b), (c), and (e) which 
are relevant to the dynamics of accelerated, matter, and radiation 
eras, respectively.
\begin{itemize}
\item Accelerated point (b) 
\be
x_c=\frac{\lambda}{\sqrt{6}q_s}\,,\quad 
\tilde{y}_c=\sqrt{1-\frac{\lambda^2}{6q_s}}\,,\quad
\Omega_b=0\,,\quad \Omega_r=0\,,\quad 
\Omega_{\phi}=1\,,\quad
w_{\phi}=w_{\rm eff}=-1+\frac{\lambda^2}{3q_s}
\label{pointb}\,.
\ee
\item $\phi$MDE point (c) 
\be
x_c=-\frac{\sqrt{6}Q}{3 q_s}\,,\quad 
\tilde{y}_c=0\,,\quad
\Omega_b=0\,,\quad \Omega_r=0\,,\quad 
\Omega_{\phi}=w_{\rm eff}=\frac{2Q^2}{3q_s}\,,\quad
w_{\phi}=1\,.
\label{pointc}
\ee
\item Radiation point (e) 
\be
x_c=0\,,\quad \tilde{y}_c=0\,,\quad
\Omega_b=0\,,\quad \Omega_r=1\,,\quad 
\Omega_{\phi}=0\,,\quad 
w_{\rm eff}=\frac{1}{3}\,.
\label{pointe}
\ee
\end{itemize}
The point (b) can drive a late-time cosmic acceleration 
under the condition $w_{\rm eff}<-1/3$. 
This translates to 
\be
\lambda^2<2q_s\,.
\label{cond1}
\ee
{}From Eq.~(\ref{weffge}), there is a possibility that 
the scaling solution (a) leads to 
the acceleration for $|Q| \gg |\lambda|$. 
As shown in Ref.~\cite{Amendola:2006qi}, however, 
such a large coupling $|Q|$ is 
hardly compatible with the existence of $\phi$MDE (c) 
with $\Omega_{\phi} \ll 1$. 
As we will also see below, if point (b) is stable, then 
point (a) is not. Hence we will focus on the case in which 
$\phi$MDE is followed by point (b).

Besides point (e), there exist the scaling fixed points (f) and (g) 
given by 
\ba
& &
{\rm (f)}~~~
x_c=\frac{2\sqrt{6}}{3\lambda}\,,\quad 
\tilde{y}_c=\sqrt{\frac{4q_s}{3\lambda^2}}\,,\quad
\Omega_b=0\,,\quad \Omega_r=1-\frac{4q_s}{\lambda^2}\,,
\quad \Omega_{\phi}=\frac{4q_s}{\lambda^2}\,,\quad
w_{\phi}=w_{\rm eff}=\frac{1}{3}\,,\\
& &
{\rm (g)}~~~x_c=-\frac{1}{\sqrt{6}Q}\,,\quad \tilde{y}_c=0\,,\quad
\Omega_b=0\,,\quad \Omega_r=1-\frac {q_s}{2Q^2}\,,\quad 
\Omega_{\phi}=\frac{q_s}{6 Q^2}\,,\quad 
w_{\phi}=1\,,\quad
w_{\rm eff}=\frac{1}{3}\,, 
\ea
both of which can be potentially used for the 
radiation era.
For point (f), however, $\Omega_r$ is 
negative under the condition (\ref{cond1}). 
The point (g) can be responsible for the radiation era 
only for $Q^2 \gg q_s$, which was exploited for the generation 
of primordial dark matter halos in Ref.~\cite{Savastano:2019zpr}.
Unless some screening of fifth forces occurs after radiation-matter 
equality, point (g) is not followed by $\phi$MDE with 
$\Omega_\phi \ll 1$. Hence we use neither (f) nor (g)  
for the fixed point of radiation era in this paper. 
In other words, we consider the cosmological sequence of fixed points:  
(e) $\to$ (c) $\to$ (b).

The stability of fixed points is established by perturbing 
Eqs.~(\ref{auto1})-(\ref{auto4}) with the
linear perturbations $\delta x$, $\delta \tilde{y}$, 
$\delta \Omega_b$, and $\delta \Omega_r$.
The signs of eigenvalues of $4 \times 4$ Jacobian matrices for 
these perturbations determine whether the fixed points 
are stable or not \cite{CLW,CST06}.
For points (e) and (c), some of the eigenvalues are positive, so 
they are not stable nodes. 
In other words, the $\phi$MDE (c), which is preceded by point (e), 
should eventually come to end to realize cosmic acceleration. 
Provided that the condition (\ref{cond1}) 
is satisfied, three of the eigenvalues for point (b) are negative. 
The other eigenvalue is negative for 
\be
\lambda (\lambda+Q)<3q_s\,,
\label{cond2}
\ee
under which point (b) is stable. 
On the other hand, the stability of point (a) requires that 
$\lambda (\lambda+Q)>3q_s$, which is opposite to the inequality (\ref{cond2}). 
This means that, under the condition (\ref{cond2}), the $\phi$MDE is 
followed by point (b) instead of point (a). 
We show a phase-space plot in Fig.~\ref{fig3} of Appendix 
to confirm the attractor property of point (b).

\begin{figure}[H]
\begin{center}
\includegraphics[height=3.4in,width=3.5in]{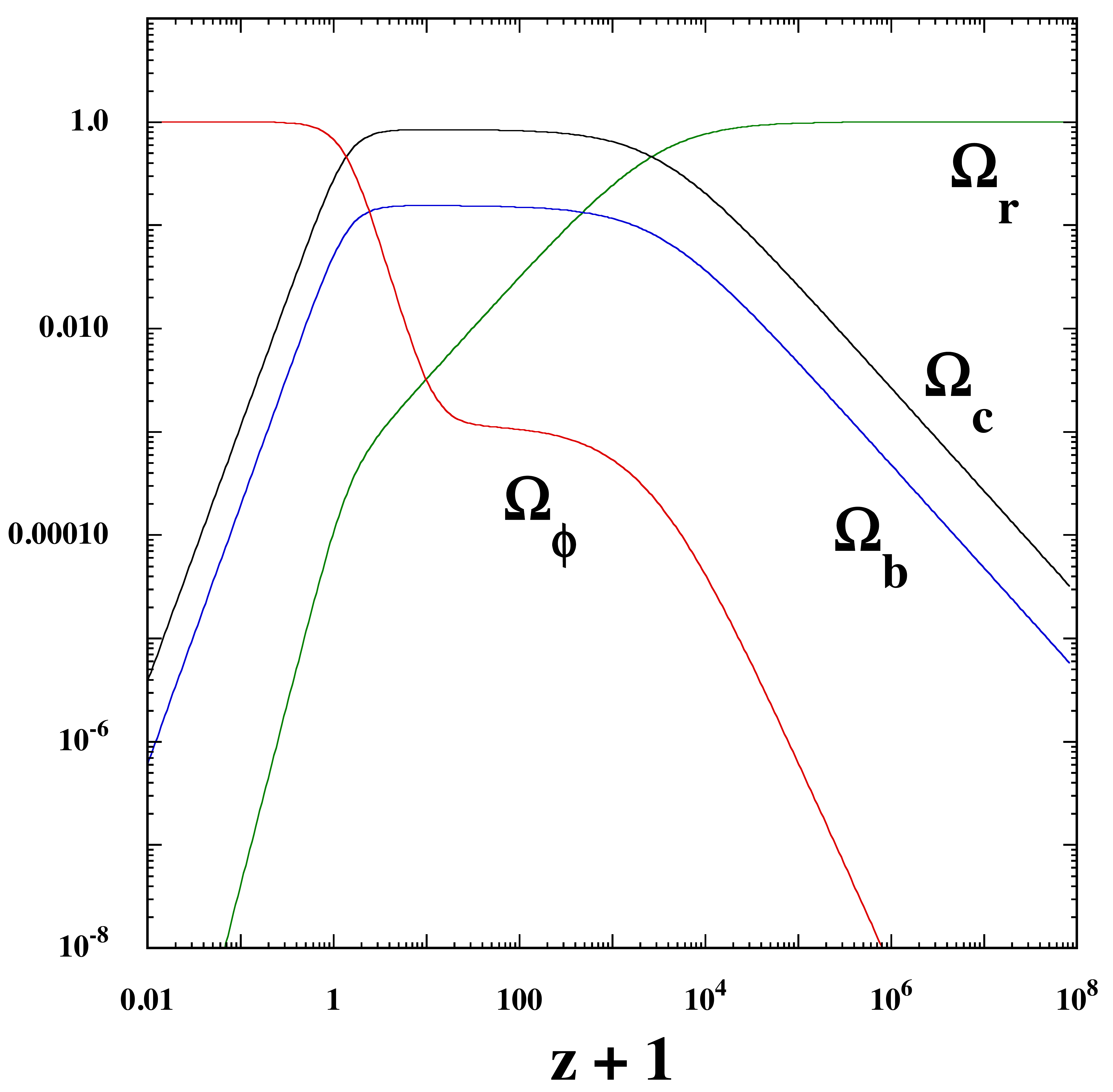}
\includegraphics[height=3.4in,width=3.5in]{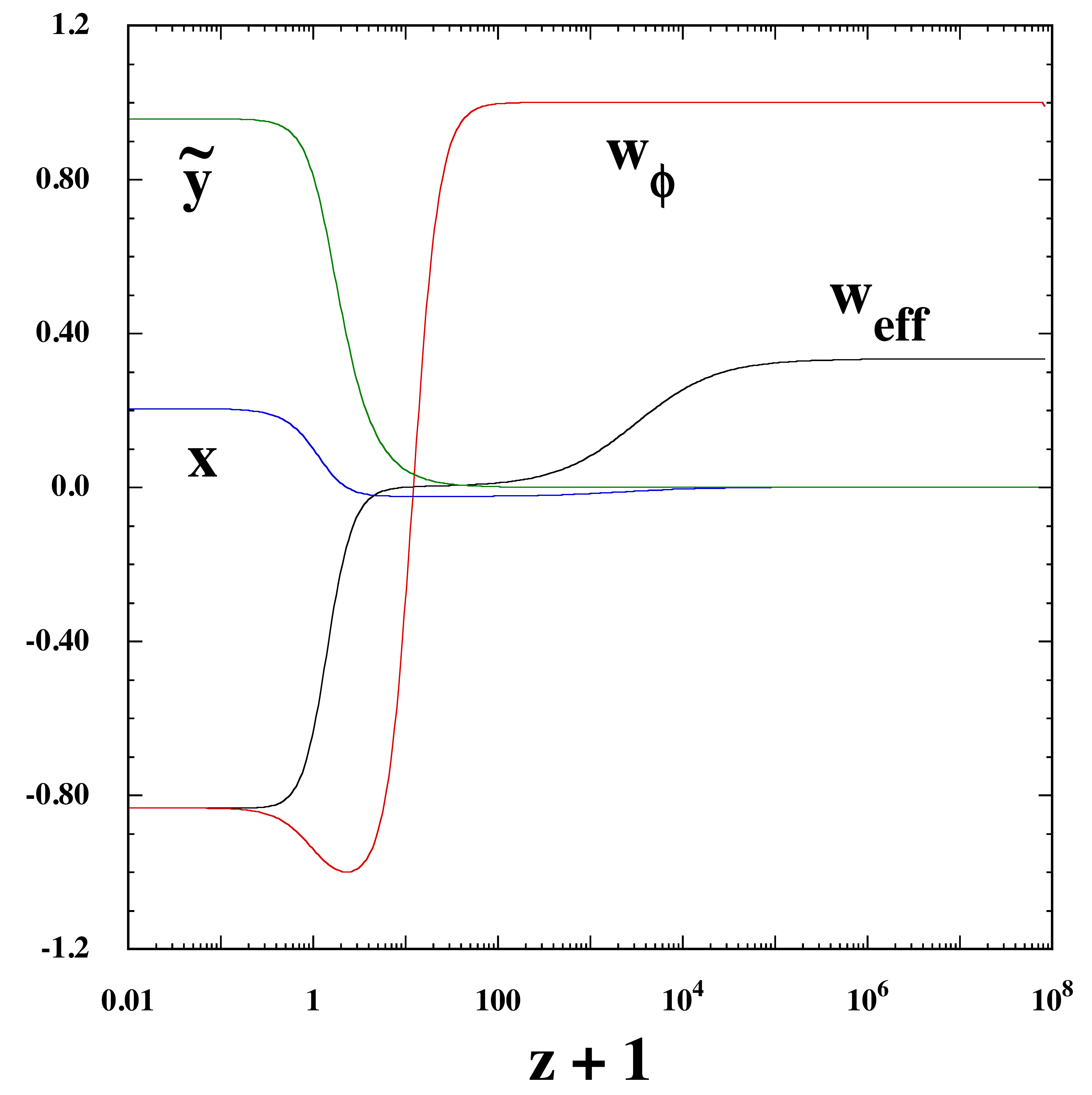}
\end{center}
\caption{\label{fig1} 
(Left)
Evolution of $\Omega_{\phi}$, $\Omega_c$, $\Omega_b$, 
and $\Omega_r$ versus $z+1$ for $\lambda=1$, $Q=0.07$, 
and $\beta=0.5$, where $z=a_0/a-1$ and $a_0$ is today's value of $a$.
The initial conditions are chosen to be
$x=1.0 \times 10^{-13}$, 
$\tilde{y}=1.0 \times 10^{-14}$, 
$\Omega_b=5.8 \times 10^{-6}$, and 
$\Omega_r=0.999962$ at redshift 
$z=8.3 \times 10^7$. 
(Right)
Evolution of $x$, $\tilde{y}$, $w_{\phi}$, 
and $w_{\rm eff}$ for the same model parameters 
and initial conditions as those used in the left. 
}
\end{figure}

In Fig.~\ref{fig1}, we plot the evolution 
of $\Omega_{\phi}$, $\Omega_c$, $\Omega_b$, and $\Omega_r$ (left) 
and $x$, $\tilde{y}$, $w_{\phi}$, and $w_{\rm eff}$ (right)
for the model parameters $\lambda=1$, $Q=0.07$, and $\beta=0.5$.
We observe that the radiation fixed point (e) is followed 
by the $\phi$MDE (c) with nearly constant values 
$\Omega_{\phi}=w_{\rm eff} \simeq 2Q^2/(3q_s)$. 
The field equation of state $w_{\phi}$ during the $\phi$MDE 
is close to 1, whose property is attributed to the kinetically 
driven evolution satisfying $|x| \simeq \sqrt{6}|Q|/(3 q_s) \gg \ty$.

Since the model parameters $\lambda$ and $\beta$ used in Fig.~\ref{fig1}
satisfy the two conditions (\ref{cond1}) and (\ref{cond2}), the solutions 
finally approach the stable accelerated point (b) 
with the asymptotic values $x=0.204$, $\tilde{y}=0.957$, and 
$w_{\phi}=w_{\rm eff}=-0.833$. They are in good agreement 
with the numerical results shown in the right panel. 
In this case, we observe that $w_{\phi}$ temporarily reaches the 
minimum around $-1$ at redshift $z \simeq 1.3$ and 
then it increases toward the asymptotic value $-0.833$. 
This evolution of $w_{\phi}$ is different from that for the model 
with $Q=0$ where $w_{\phi}$ is close to $-1$ 
during the matter era and 
finally approaches $-1+\lambda^2/(3q_s)$ \cite{Kase:2019mox}. 
Thus, the two models can be observationally distinguished 
from each other.

We note that the values of $\Omega_{\phi}$ and $w_{\rm eff}$ on the $\phi$MDE are reduced by 
the positive coupling $\beta$. 
The presence of positive $\beta$ also leads to the decreases of $w_{\phi}$ and $w_{\rm eff}$
on the attractor point (b). 
For increasing $\beta$, the asymptotic value of $w_{\phi}$ gets closer to $-1$.  
Nevertheless, the early evolution of $w_{\phi}$ is different from that 
in the $\Lambda$CDM model. It remains to be seen whether the problem 
of $H_0$ tension in the $\Lambda$CDM model can be alleviated in 
our interacting model with $Q \neq 0$ and $\beta \neq 0$.

\section{Perturbation dynamics for a model with $\phi$MDE}
\label{persec}

In this section, we study the evolution of linear 
scalar perturbations 
for the theory (\ref{action}) with the interacting 
Lagrangian (\ref{Lagcon}). 
On the flat FLRW background, we consider 
the perturbed line element given by 
\be
{\rm d}s^2=-(1+2\alpha) {\rm d}t^2
+2 \partial_i \chi {\rm d}t {\rm d}x^i
+a^2(t) \left[ (1+2\zeta) \delta_{ij}
+2\partial_i \partial_j E \right] {\rm d}x^i {\rm d}x^j\,.
\label{permet}
\ee
The quantities $\alpha$, $\chi$, $\zeta$, and $E$ 
are scalar perturbations, which  
depend on both $t$ and spatial coordinate $x^i$. 
Here and in the following, we use
the notation 
$\partial_i \chi=\partial \chi/\partial x^i$.
We do not take the tensor perturbation into account in 
Eq.~(\ref{permet}), but it propagates in the same manner 
as in standard general relativity. 
We decompose the scalar field $\phi$ into the background 
part $\bar{\phi}(t)$ and the perturbation 
$\delta \phi$, as 
\be
\phi=\bar{\phi}(t)+\delta \phi (t,x^i)\,,
\label{phide}
\ee
where we omit the bar from background quantities in the following.

For perfect fluids, we take CDM, baryons, and radiation into account 
labelled by $I=c, b, r$, respectively.
{}From Eq.~(\ref{Jmure}), it follows that the number 
${\cal N}_I=n_I  a^3$ of each fluid is conserved at 
the background level.  
We express the temporal and spatial components of 
$J_I^{\mu}$, respectively,  as, 
\be
J_I^0={\cal N}_I+\delta J_I\,,\qquad 
J_I^i=\frac{1}{a^2(t)} \delta^{ik} \partial_k \delta j_I\,,
\label{JI}
\ee
where $\delta J_I$ and $\delta j_I$ are the scalar perturbations.
The scalar velocity potential $v_I$ is related to the spatial component 
of four velocity $u_{Ii}=J_{Ii}/(n_I\sqrt{-g})$, as
\be
u_{Ii}=-\partial_{i}v_I\,.
\ee
Since $J_{Ii}=J_I^0 g_{0i}+J_I^j g_{ij}
={\cal N}_I \partial_{i} \chi+
\partial_{i} \delta j_I$ at linear 
order in perturbations, it follows that 
\be
\partial_i \delta j_I=-{\cal N}_I \left( \partial_i \chi 
+\partial_i v_I \right)\,.
\label{delj}
\ee

On using Eq.~(\ref{lc}), there is also the following relation 
\be
\ell_c=-\int^{t} \left[ 1+f_1(\tilde{t}) \right] 
\rho_{c,n_c} (\tilde{t})\,{\rm d} \tilde{t}
-\left( 1+f_1 \right) \rho_{c,n_c}v_c
+\frac{f_{1,Z}\rho_c-f_{2,Z}}{n_c} 
\left( \dot{\phi}\,v_c-\delta \phi \right)\,,
\ee
up to first order in perturbations.
On the other hand, the Lagrange multipliers $\ell_I$ 
for baryons and radiation are
\be
\ell_I=-\int^{t} 
\rho_{I,n_I} (\tilde{t})\,{\rm d} \tilde{t}
-\rho_{I,n_I}v_I\,,\qquad
{\rm with}~I=b,r\,.
\ee
The fluid density is given by 
$\rho_I=\rho_I(t)+\delta \rho_I (t,x^i)$, 
where the perturbed part is 
\be
\delta \rho_I=\frac{\rho_{I,n_I}}{a^3} 
\left[ \delta J_I-{\cal N}_I \left( 3\zeta+\partial^2 E 
\right) \right]\,,
\ee
where $\partial^2 E=\sum_{i=1}^3 \partial_i^2 E$.
In this case, we have  
\be
\delta n_I= \frac{\delta \rho_I}{\rho_{I,n_I}}
-\frac{{\cal N}_I (\partial v_I)^2}{2a^5}
-(3\zeta+\partial^2E)\frac{\delta \rho_I}{\rho_{I,n_I}}
-\frac{{\cal N}_I(\zeta+\partial^2E)(3\zeta-\partial^2E)}{2a^3}
+{\cal O}(\varepsilon^3)\,,
\label{deln}
\ee
where $(\partial v_I)^2=\sum_{i=1}^3 (\partial_i v_I)^2$, and
$\varepsilon^n$ represents the $n$-th order of perturbations.
We also introduce the sound speed squared of each matter 
fluid, as 
\be
c_I^2=\frac{n_I \rho_{I,n_I n_I}}{\rho_{I,n_I}}\,.
\ee
The quantities $\rho_I(n_I)$, $X$, and $Z$, when are expanded 
up to second order in perturbations, are given by 
\ba
\rho_I (n_I)
&=&\rho_I+\left( \rho_I+P_I \right) 
\frac{\delta n_I}{n_I}+\frac{1}{2} \left( \rho_I+P_I \right) 
c_I^2 \left( \frac{\delta n_I}{n_I} \right)^2
+{\cal O} (\varepsilon^3)\,,\\
X &=& \frac{1}{2} \dot{\phi}^2+\dot{\phi} (\dot{\delta \phi}-\dot{\phi} \alpha)
+\frac{1}{2} \left[  (\dot{\delta \phi}-2\dot{\phi} \alpha)^2
-\frac{1}{a^2}(\partial \delta \phi+\dot{\phi} \partial \chi)^2
\right]+{\cal O} (\varepsilon^3)\,,
\label{deltaX}\\
Z &=& \dot{\phi}+\dot{\delta \phi}-\dot{\phi} \alpha
+\frac{1}{2a^2} \left[ \dot{\phi} \left\{ 3a^2 \alpha^2
-(\partial_i \chi)^2+(\partial_i v_c)^2 \right\}-2a^2 \alpha \dot{\delta \phi}
-2\partial_i \delta \phi (\partial_i \chi+\partial_i v_c) 
\right] +{\cal O} (\varepsilon^3)\,.
\label{delZ}
\ea
In the following, we consider the case in which the equations 
of state and sound speed squares of both CDM and baryons vanish, i.e., 
\be
w_c^2=0\,,\qquad w_b^2=0\,,\qquad 
c_c^2=0\,,\qquad c_b^2=0\,.
\ee

As in Refs.~\cite{Kase:2019veo,Kase:2019mox}, 
the linear perturbation equations of motion follow by expanding 
the action (\ref{action}) up to quadratic order.
Alternatively, we can also derive them by using the covariant 
equations of motion (\ref{Ein}), (\ref{Tcon}), and 
(\ref{Tcon2}) at first order.
The explicit form of second-order action will be presented 
for a more general interacting Lagrangian $L(n_c, \phi, X, Z)$ 
as a future work \cite{KTfuture}. 
Here, we show the perturbation equations for the interacting 
Lagrangian (\ref{Lagcon}) in a gauge-ready form. 
For this purpose, we introduce the following 
gauge-invariant quantities,
\ba
&&
\delta \phi_{\rm N}=\delta \phi+\dot{\phi}\left(\chi-a^2 \dot{E}\right)\,,
\qquad 
\delta \rho_{I\rm N}=\delta \rho_I+\dot{\rho}_I \left(\chi-a^2 \dot{E}\right)\,,
\qquad 
v_{I{\rm N}}=v_I+\chi-a^2 \dot{E}\,,\nonumber \\
&&
\Psi=\alpha+\frac{{\rm d}}{{\rm d}t} 
\left( \chi - a^2 \dot{E} \right)\,,\qquad 
\Phi=\zeta+H \left( \chi - a^2 \dot{E} \right)\,,
\ea
together with the dimensionless variables, 
\be
\delta_{I{\rm N}}= \frac{\delta \rho_{I\rm N}}{\rho_I}\,,\qquad
\delta \varphi_{\rm N}=\frac{H}{\tp}\delta\phi_{\rm N}\,,\qquad
V_{I\rm N}=H v_{I{\rm N}}\,,\qquad 
{\cal K}=\frac{k}{aH}\,,
\ee
where $k$ is a comoving wavenumber.

In Fourier space, the full set of linear perturbation 
equations of motion is then given by 
\ba
& & 
6 q_s x^2 \delta \varphi_{\rm N}'-6 \Phi'+6 \left( 
1-q_s x^2 \right) \left( \xi \delta \varphi_{\rm N} 
+\Psi \right)-2{\cal K}^2 \Phi
+3 \left( 3\Omega_c+3\Omega_b+4\Omega_r 
\right) \delta \varphi_{\rm N}  \nonumber \\
& &
+3 \left( \Omega_c \delta_{c{\rm N}} 
+\Omega_b \delta_{b{\rm N}}
+\Omega_r \delta_{r{\rm N}}  \right)=0\,,
\label{pereq1}\\
& &
\Phi'-\Psi-\xi \delta \varphi_{\rm N}
+\frac{3}{2} \left( \Omega_c+2m \beta  x^2 
\right) \left( V_{c{\rm N}}-\delta \varphi_{\rm N} \right)
+\frac{3}{2} \Omega_b 
\left( V_{b{\rm N}}-\delta \varphi_{\rm N} \right) 
\nonumber \\
& &
+2\Omega_r 
\left( V_{r{\rm N}}-\delta \varphi_{\rm N} \right)=0\,,\\
& &
\delta'_{I{\rm N}}+3 \left( c_I^2-w_I \right) \delta_{I{\rm N}}
+\left( 1+w_I \right) \left( {\cal K}^2 V_{I{\rm N}}
+3\Phi' \right)=0\,,\qquad ({\rm for}~I=c,b,r), 
\label{pereq3}\\
& & 
\left( \Omega_c+2m \beta x^2 \right) V'_{c{\rm N}}
-\left[ \xi \left( \Omega_c+2m \beta  x^2 \right)
-2m \beta x^2 (3+2\epsilon_{\phi}) 
-\sqrt{6} Q x \Omega_c
\right] V_{c{\rm N}}
-\Omega_c \Psi \nonumber \\
& &
-2m \beta x^2 \delta \varphi'_{\rm N}
+\left[ 2m \beta x ( \xi -3-2\epsilon_{\phi})
-\sqrt{6}Q \Omega_c \right] x
\delta \varphi_{\rm N} =0\,,
\label{pereq4}\\
& &
V'_{I{\rm N}}-\left( \xi+3 c_I^2 \right) V_{I{\rm N}} 
-\Psi-\frac{c_I^2}{1+w_I} \delta_{I{\rm N}}=0\,,
\qquad ({\rm for}~I=b,r), \\
& &
\delta \varphi''_{\rm N}+ \left( 3-\xi +2\epsilon_{\phi} 
\right)\delta \varphi'_{\rm N}
+\left[ \hat{c}_s^2 {\cal K}^2-\xi'-3\xi+\epsilon'_{\phi}
+\epsilon_{\phi}^2
+(3-\xi)\epsilon_{\phi}
+\frac{3}{q_s} \left( \lambda^2 \tilde{y}^2
+Q^2 \Omega_c \right) \right] 
\delta \varphi_{\rm N} \nonumber \\
& &+3\hat{c}_s^2 \Phi'-\Psi'-2\left(3+\epsilon_{\phi}\right)\Psi
-\frac{m \beta}{q_s} \delta'_{c{\rm N}}
+\frac{\sqrt{6} Q \Omega_c}{2q_s x} \delta_{c{\rm N}}=0\,,
\label{delphi} \\
& &
\Psi=-\Phi\,,
\label{pereq6}
\ea
where we remind that the prime denotes the derivative 
with respect to $N=\ln a$, and 
\be
\hat{c}_s^2=1-\frac{m \beta}{q_s}\,.
\ee

Equations (\ref{pereq1})-(\ref{pereq6}) are written in the gauge-ready 
form, in that they can be used for any gauge choices of interest.
For example, if we choose the unitary gauge characterized by 
$\delta \phi=0$ and $E=0$, the dynamical perturbations 
correspond to the curvature perturbation 
${\cal R}=\Phi-\delta \varphi_{\rm N}$ and the 
density perturbations 
$\delta \rho_{I{\rm u}}=\delta \rho_{I{\rm N}}
-\dot{\rho}_I \delta \phi_{\rm N}/\dot{\phi}$. 
We can eliminate nondynamical variables like $\alpha$, 
$\chi$, and $v_I$ from the second-order action 
by using Eqs.~(\ref{pereq1})-(\ref{pereq3}). 
Following a procedure similar to that performed 
in Refs.~\cite{Kase:2019veo,Kase:2019mox} for perturbations 
deep inside the Hubble radius, 
there are neither ghosts nor Laplacian instabilities 
for dynamical perturbations ${\cal R}$ and 
$\delta \rho_{c{\rm u}}$ under the conditions,
\ba
q_s &=&1+2\beta>0\,,
\label{thecon1} \\
q_c &=& 1+\frac{2m \beta x^2}{\Omega_c}
>0\,,\label{thecon2}\\
c_s^2 &=& \hat{c}_s^2+\frac{2 \beta^2 m^2 x^2}
{q_s (2 \beta m x^2+\Omega_c)} \geq 0\,.
\label{thecon3}
\ea
In the limit that $c_c^2 \to 0$, the effective sound speed squared 
of CDM vanishes, so that there is no additional pressure 
modifying the evolution of CDM density perturbations. 
The effective sound speed squared of the scalar field 
in the small-scale limit corresponds to $c_s^2$ given in 
Eq.~(\ref{thecon3}). In comparison to the value $\hat{c}_s^2$, there is a 
correction to $c_s^2$ arising from a kinetic mixing between 
the scalar field and CDM \cite{Kase:2019veo,Kase:2019mox}.
This correction term is positive under the no-ghost 
conditions (\ref{thecon1}) and (\ref{thecon2}).
Provided that $\hat{c}_s^2 \geq 0$, the positivity of 
$c_s^2$ is always ensured.

In the following, we study the evolution of perturbations 
after the onset of $\phi$MDE. For this purpose, 
we ignore the contribution of radiation perturbations 
to Eqs.~(\ref{pereq1})-(\ref{pereq6}) and set $\Omega_r=0$. 
For CDM and baryons, Eq.~(\ref{pereq3}) reduces to 
\be
\delta_{I{\rm N}}'+{\cal K}^2 V_{I{\rm N}}+3\Phi'=0\,,
\label{delIN}
\ee
where $I=c,b$.
The CDM velocity potential $V_{c{\rm N}}$ satisfies the first-order 
differential Eq.~(\ref{pereq4}), 
while the baryon velocity potential $V_{b{\rm N}}$ obeys
\be
V_{b{\rm N}}'- \xi V_{b{\rm N}} -\Psi=0\,.
\label{pereq5}
\ee
Differentiating Eq.~(\ref{delIN}) with respect to $N$ and using 
Eqs.~(\ref{pereq4}) and (\ref{pereq5}), it follows that 
\ba
& & \delta_{c{\rm N}}''+\nu_1 \delta_{c{\rm N}}'
+\nu_2 {\cal K}^2 +3\Phi''+3 \nu_1\Phi'=0\,,
\label{delceq} \\
& & \delta_{b{\rm N}}''+\left( 2+\xi \right) \delta_{b{\rm N}}'
+{\cal K}^2 \Psi+3\Phi''+3 \left( 2+\xi \right) \Phi'=0\,,
\label{delbeq}
\ea
where 
\ba
\nu_1 &=& 2+\xi+\frac{[2m \beta (3+2\epsilon_{\phi})x
+\sqrt{6} Q \Omega_c]x}{\Omega_c+2m \beta x^2}\,,\\
\nu_2 &=& \frac{\Omega_c \Psi+2m \beta x^2 \delta \varphi_{\rm N}'
+[2m \beta (3-\xi+2\epsilon_{\phi})x
+\sqrt{6}Q \Omega_c] x\delta \varphi_{\rm N}
}{\Omega_c+2m \beta x^2}\,.
\ea

Now, we employ the quasi-static approximation for perturbations 
deep inside the sound horizon \cite{Boisseau,DKT11}. 
Since the dominant contributions to the perturbation equations 
are those containing ${\cal K}^2$, $\delta_{c{\rm N}}$, 
$\delta'_{c{\rm N}}$, and $\delta_{b{\rm N}}$, 
Eqs.~(\ref{pereq1}), (\ref{delphi}), and (\ref{pereq6}) give the following relations, 
\ba
& &
\Psi=-\Phi \simeq -\frac{3}{2{\cal K}^2} 
\left(  \Omega_c \delta_{c{\rm N}} 
+\Omega_b \delta_{b{\rm N}} \right)\,,
\label{quasi1} \\
& &
\delta \varphi_{\rm N} \simeq \frac{1}{q_s \hat{c}_s^2 {\cal K}^2} 
\left( m \beta \delta'_{c{\rm N}}-\frac{\sqrt{6} Q \Omega_c}{2x}
\delta_{c{\rm N}} \right)\,.
\label{quasi2}
\ea
Under the quasi-static approximation, we can ignore the terms 
$3\Phi''$, $3 \nu_1\Phi'$, and $3 \left( 2+\xi \right) \Phi'$ 
in Eqs.~(\ref{delceq}) and (\ref{delbeq}) relative to the others. 
Substituting Eqs.~(\ref{quasi1}), (\ref{quasi2}) and the $N$ derivative 
of Eq.~(\ref{quasi2}) into Eqs.~(\ref{delceq}) and (\ref{delbeq}), it follows that 
\ba
& &
\delta_{c{\rm N}}''+\nu \delta_{c{\rm N}}'
-\frac{3}{2G} \left( G_{cc} \Omega_c \delta_{c{\rm N}}
+G_{cb} \Omega_b \delta_{b{\rm N}} \right) \simeq 0\,,
\label{delceq2} \\
& &
\delta_{b{\rm N}}''+\left( 2+\xi \right) \delta_{b{\rm N}}'
-\frac{3}{2G} \left( G_{bc} \Omega_c \delta_{c{\rm N}}
+G_{bb} \Omega_b \delta_{b{\rm N}} \right) \simeq 0\,,
\label{delbeq2}
\ea
where 
\be
\nu=\frac{2m \beta (1+2\beta)(5+\xi+2\epsilon_{\phi})x^2 
+(2+\xi+\sqrt{6}Qx)[1+(2-m)\beta]\Omega_c}
{2m \beta (1+2\beta) x^2+\Omega_c 
[1+(2-m)\beta]}\,,
\ee
and 
\ba
& &
G_{cc} = \frac{1+r_1}{1+r_2}G\,,\qquad 
G_{cb} = \frac{1}{1+r_2}G\,,\label{GCDM} \\
& &
G_{bc} = G_{bb}=G\,,\label{Gba}
\ea
with
\ba
r_1 &=& \frac{2Q[3Q \Omega_c+\sqrt{6} m \beta x 
(2+\epsilon_{\phi}+\sqrt{6}Q x)]}
{3\Omega_c [1+(2-m)\beta]}\,,\\
r_2 &=& \frac{2m \beta (1+2\beta)x^2}
{\Omega_c [1+(2-m)\beta]}\,.
\ea
In the limit that $\beta \to 0$, we have $r_1=2Q^2$ and $r_2=0$, 
so that $G_{cc}=(1+2Q^2)G$ and $G_{cb}=G_{bc}=G_{bb}=G$. 
Since $G_{cc}$ is larger than $G$, the growth rate 
of $\delta_{c{\rm N}}$ 
is larger than that for $Q=0$. 
In the other limit $Q \to 0$, we have $r_1=0$ and hence 
$G_{cc}=G_{cb}=G/(1+r_2)$ and $G_{bc}=G_{bb}=G$.  
In this case, under the conditions (\ref{thecon1})-(\ref{thecon3}), 
both $G_{cc}$ and $G_{cb}$ are smaller than $G$ 
for $m \beta>0$.

For the theories with $\beta \neq 0$ and $Q \neq0$, 
$G_{cc}$ can be either smaller or larger than $G$
depending on the model parameters.
We note that $G_{cb}$ is equivalent to $G_{cc}$ in the limit $Q \to 0$.

During the $\phi$MDE, $G_{cc}$ and $G_{cb}$ are given, respectively, by 
\ba
G_{cc}
&=&\left( 1+\frac{2Q^2}{1+2\beta} \right)G\,,
\label{Gcc1}\\
G_{cb}
&=& \left[ 1-\frac{4m \beta Q^2}{3-2Q^2+6(2-m)\beta^2
+\{ 12-3m+(6m -4)Q^2 \}\beta} \right]G\,.
\label{Gcb}
\ea
Under the condition (\ref{thecon1}), $G_{cc}$ is larger than $G$.

A simple approximation valid at the present time for the small 
couplings $|\beta| \ll 1$ and $|Q|\ll 1$ is given by 
\begin{equation}
G_{cc} \simeq \left( 1+2Q^2 -\frac{2\beta m x^2}{\Omega_c} \right)G\,,
\end{equation}
where we ignored the terms which are the products of $\beta$ 
and $Q$.
This shows how $G_{cc}$ can be larger or smaller than $G$ depending on the coupling parameters.

On the fixed point (b) given by Eq.~(\ref{pointb}),
$G_{cc}$ reduces to 
\be
G_{cc}=\frac{4(1+2\beta)+\lambda(2Q-\lambda)}
{1+2\beta} \frac{Q}{\lambda} G\,,
\label{Gcc2}
\ee
whereas $G_{cb}$ vanishes. 
{}From Eq.~(\ref{Gcc2}), we have $G_{cc} \to 0$ for $Q \to 0$. 
If we impose the condition $G_{cc}>0$ on the future attractor 
point (b), we require that 
$[4(1+2\beta)+\lambda(2Q-\lambda)]Q/\lambda>0$. 

Since both $G_{cc}$ and $G_{cb}$ are different from $G$, 
this affects the evolution of $\delta_{c{\rm N}}$ through Eq.~(\ref{delceq2}). 
While both $G_{bc}$ and $G_{bb}$ are equivalent to $G$, 
the modified evolution of $\delta_{c{\rm N}}$ affects 
the growth of $\delta_{b{\rm N}}$ through Eq.~(\ref{delbeq2}).
To study the evolution of total matter perturbations, 
we introduce the effective 
CDM background density $\hat{\rho}_c=(1+f_1) \rho_c=e^{Q \phi/M_{\rm pl}}\rho_c$ 
and the perturbed gauge-invariant density 
$\hat{\delta \rho}_{c{\rm N}}=e^{Q\phi/M_{\rm pl}} (\delta \rho_{c{\rm N}}+Q\rho_c 
\delta \phi_{\rm N}/M_{\rm pl})$. 
The total matter density contrast is given by 
\be
\delta_m=\frac{\hat{\delta \rho}_{c{\rm N}}+\delta \rho_{b {\rm N}}}
{\hat{\rho}_c+\rho_b}
=\left( \delta_{c{\rm N}}+\sqrt{6}Q x \delta \varphi_{\rm N} \right) 
\frac{\Omega_c}{\Omega_m}+\delta_{b{\rm N}} 
\frac{\Omega_b}{\Omega_m}\,,
\label{delmes}
\ee
where $\Omega_m=\Omega_c+\Omega_b$. 
For the perturbations deep inside the Hubble radius (${\cal K} \gg 1$), 
Eq.~(\ref{quasi2}) shows that 
the term $\sqrt{6}Q x \delta \varphi_{\rm N}$ 
in Eq.~(\ref{delmes}) 
is negligibly small relative to $\delta_{c{\rm N}}$, so  Eq.~(\ref{delmes})
reduces to 
$\delta_m \simeq \delta_{c{\rm N}} \Omega_c/\Omega_m
+\delta_{b{\rm N}} \Omega_b/\Omega_m$. 
The quantity related to the measurement of redshift-space distortions 
is $f (z) \sigma_8 (z)$, where $f=\delta_m'/\delta_m$ 
is the growth rate of 
matter perturbations that depends on the redshift $z$.
On the critical points, $f$ is constant and can be obtained  analytically, 
since all the coefficients of Eqs.~(\ref{delceq2}) and 
(\ref{delbeq2}) are constant.  
If we neglect the contribution of baryons to Eq.~(\ref{delceq}),  
the value of $f$ corresponding to 
the growing-mode solution is given by 
\begin{equation}
    f= \frac{1}{2} \left( \sqrt{\nu^2+6\frac{G_{cc}}{G} \Omega_c}
     -\nu\right)\,.
\end{equation} 
For instance, $f=1+2Q^2/q_s$ during the $\phi$MDE. 
The matter growth rate is always larger than 
in standard general relativity ($f=1$),
but, for $\beta>0$, it is smaller than in a pure 
energy-exchange model with $\beta=0$ and 
$Q \neq 0$. In general, however, we need the numerical integration to know the 
precise evolution of $\delta_m$ at low redshifts, which we describe next.

\begin{figure}[H]
\begin{center}
\includegraphics[height=3.4in,width=3.5in]{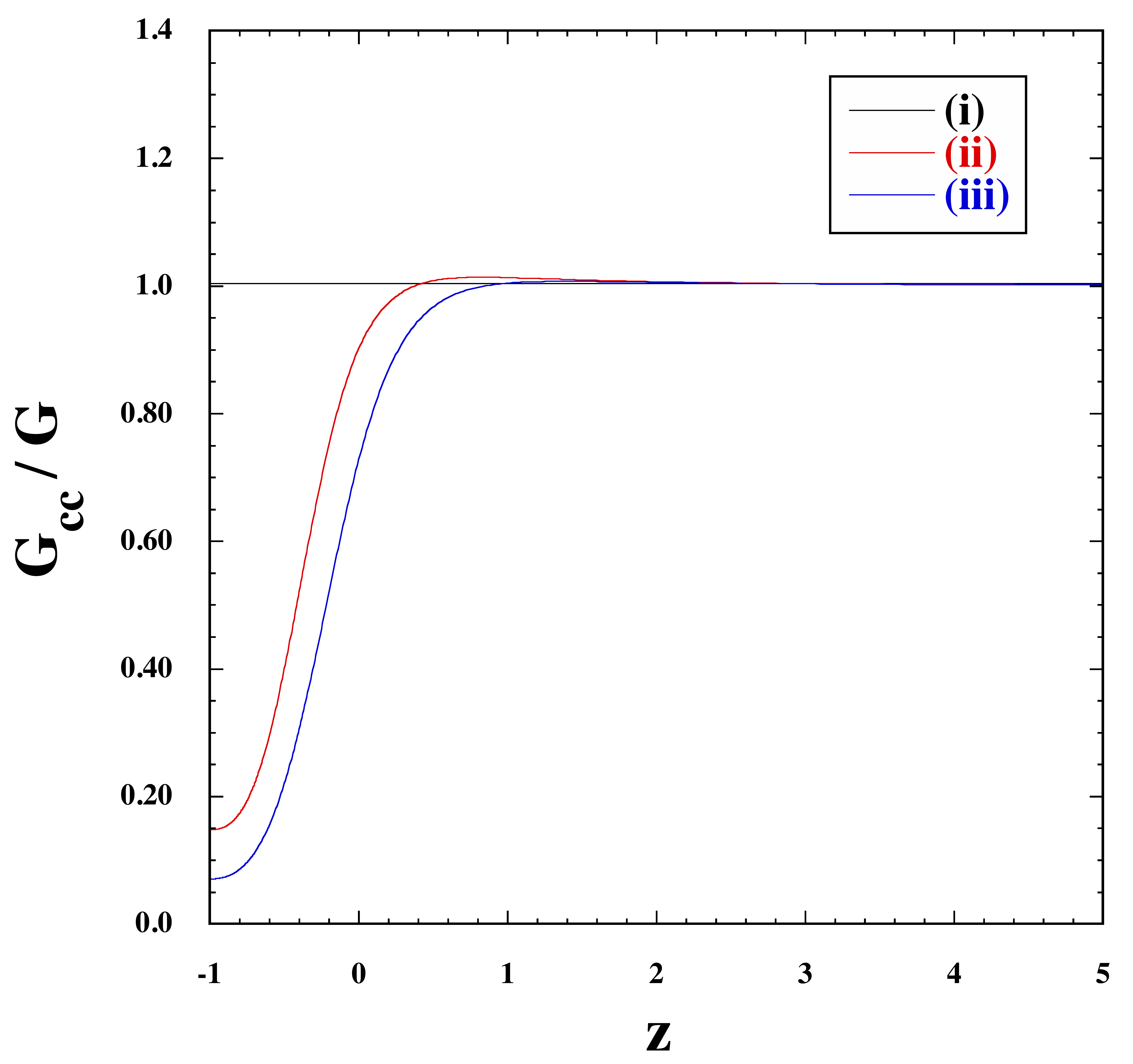}
\includegraphics[height=3.4in,width=3.5in]{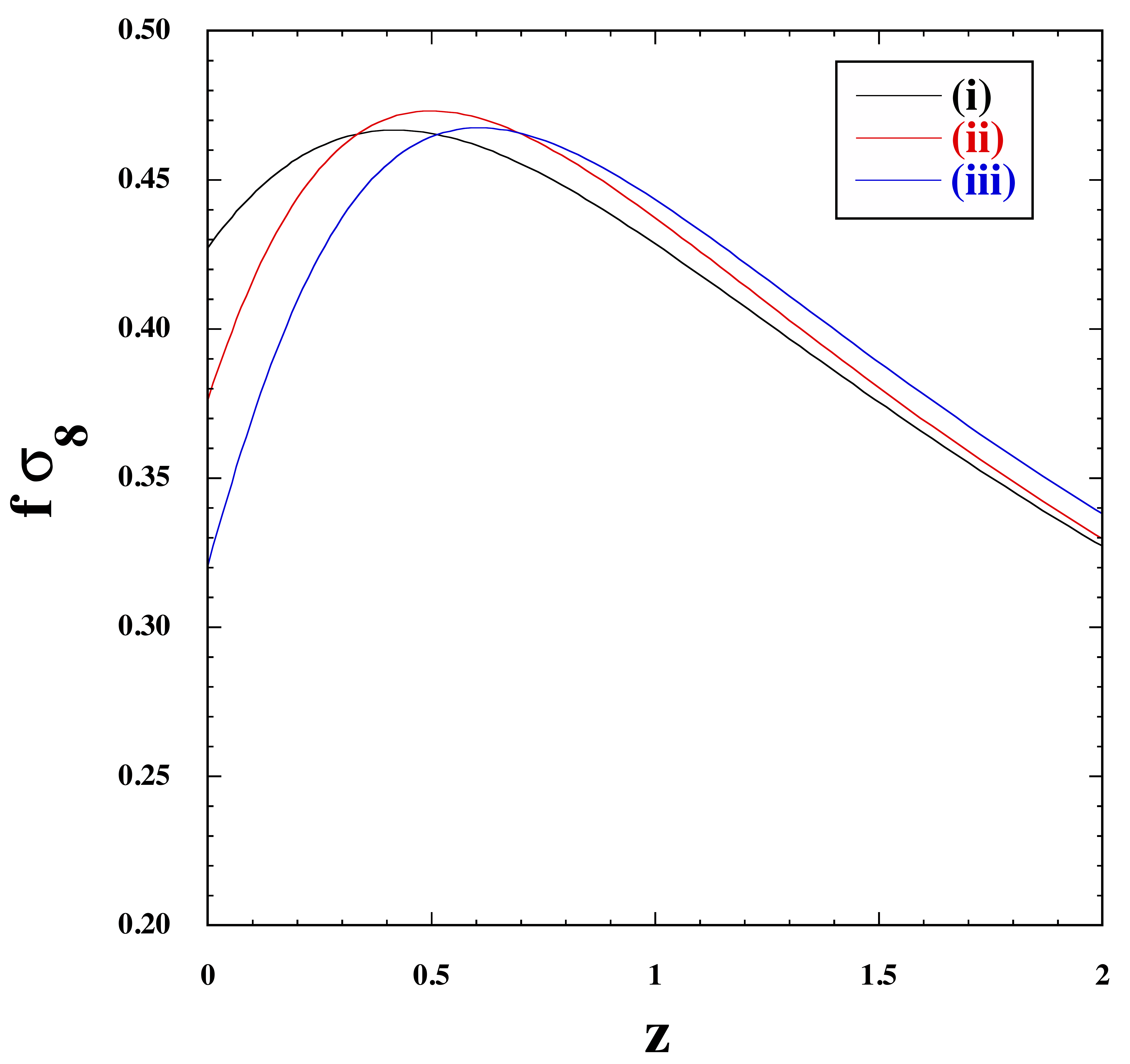}
\end{center}
\caption{\label{fig2} 
(Left) 
Evolution of $G_{cc}/G$ versus the redshift $z$ 
for $\lambda=1$ in three different cases: 
(i) $Q=0.04$, $\beta=0$, $m=2$, 
(ii) $Q=0.04$, $\beta=1$, $m=2$, and 
(iii) $Q=0.02$, $\beta=0.5$, $m=3$. 
The background initial conditions are chosen 
to realize $\Omega_{\phi} \simeq 0.68$, 
$\Omega_b \simeq 0.05$, and $\Omega_r \simeq 10^{-4}$ today.
(Right) 
Evolution of $f \sigma_8$ versus the redshift $z$ 
for the three cases shown in the left panel. 
Today's values of $\sigma_8$ and ${\cal K}$ are chosen 
to be $\sigma_8 (z=0)=0.811$ and 
${\cal K}(z=0)=300$, 
respectively.
}
\end{figure}

In Fig.~\ref{fig2}, we plot the evolution of $G_{cc}/G$ (left)
and $f \sigma_8$ (right) versus the redshift $z$ for $\lambda=1$ in three different cases: 
(i) $Q=0.04$, $\beta=0$, $m=2$,
(ii) $Q=0.04$, $\beta=1$, $m=2$, and 
(iii) $Q=0.02$, $\beta=0.5$, $m=3$. 
In these cases, the background cosmology corresponds to the $\phi$MDE 
followed by the accelerated fixed point (b).
In case (i), all the terms associated with the coupling 
$\beta$ disappear from the background and 
perturbation equations, so their dynamics is 
the same as that of standard coupled quintessence 
with the exponential potential \cite{Amendola99,Amendola:2003wa}.
Indeed, $G_{cc}$ is equivalent to $(1+2Q^2)G=1.0032G$ 
in the simulation of Fig.~\ref{fig2}, with $G_{cb}=G$.

In case (ii), the analytic estimations (\ref{Gcc1})-(\ref{Gcc2}) 
show that $G_{cc}=1.0011G$ and $G_{cb}=0.9986G$
during the $\phi$MDE and $G_{cc}=0.1477G$ 
on the accelerated point (b). 
In comparison to case (i), the existence of positive coupling 
$\beta$ leads to the smaller value of $G_{cc}$ in $\phi$MDE, 
which is also confirmed numerically. 
At low redshifts, there is a transient period during which 
$G_{cc}$ gets larger than that 
in case (i). However, $G_{cc}$ starts to be smaller than $G$ 
around the redshift $z \lesssim 0.37$ and it reaches 
today's value $0.90G$. 
Thus, even for $Q \neq 0$, the presence of positive
coupling $\beta$ can realize $G_{cc}$ smaller than $G$ by today. 
In case (iii), the decrease of $G_{cc}$ at low redshifts 
is even more significant relative to case (ii). 
In this case, today's values of CDM gravitational couplings 
are $G_{cc}=0.72G$ and $G_{cb}=0.65G$
with the future asymptotic value $G_{cc}=0.070G$. 

As we observe in the right panel of Fig.~\ref{fig2}, the decrease of 
$G_{cc}$ less than $G$, which occurs in cases (ii) and (iii) at low redshifts, 
leads to  values of $f \sigma_8$ smaller than that in case (i). 
This suppression of growth of $\delta_m$ is the consequence of momentum transfer 
associated with the positive coupling $\beta$. 
The evolution of $f \sigma_8$ depends on the couplings 
$Q$, $\beta$ as well as the model parameters $m$, $\lambda$. 
For $Q$ close to 0, $G_{cc}$ approaches $G_{cb}=G/(1+r_2)$, 
which is in the range $G_{cb}<G$ for $m \beta>0$. 
In such cases, it is easy to realize a CDM gravitational interaction 
smaller than the Newton gravitational constant. 
The nonvanishing coupling $Q$ works to enhance $G_{cc}$, 
but the energy transfer is required for 
the existence of $\phi$MDE.

We note that our interacting model with the $\phi$MDE
is different from the early dark energy recently studied in Ref.~\cite{Hill}, 
in that the latter only modifies the background dynamics with 
the standard growth of matter perturbations. 
In our model, the weak cosmic growth rate is realized by 
the momentum transfer, with the modified
early background dynamics by the energy transfer. 
Hence there is the possibility that the observational tensions of 
both $\sigma_8$ and $H_0$ are relaxed. However, clearly, we need detailed 
Markov chain Monte Carlo simulations with the recent observational data 
to see whether this is really the case or not.

\section{Conclusions}
\label{consec}

In this paper, we studied the cosmological dynamics of interacting 
theories of dark energy and dark matter by paying particular attention to 
the existence of a scaling $\phi$MDE. 
Our goal is not only to alleviate the $H_0$ tension problem by realizing 
the $\phi$MDE but also to ease the $\sigma_8$ tension problem  
by achieving a weak gravitational interaction on scales relevant to 
the growth of large-scale structures.
It is possible to satisfy these two demands by considering 
the interacting action (\ref{action0}) containing  
both energy and momentum transfers. 
The dependence of $Z$ in the Lagrangian $f_2(\phi,X,Z)$, 
where $Z=u^{\mu} \nabla_{\mu} \phi$ quantifies
the interaction between the CDM four velocity $u^{\mu}$ 
and the scalar derivative $\nabla_{\mu} \phi$, plays a crucial role 
for the realization of weak gravity.

In Sec.~\ref{scalingsec}, we derived the interacting Lagrangian 
allowing for the existence of scaling solutions which obey
the relation $\dot{\phi}/H={\rm constant}$. 
We showed that the corresponding Lagrangian is of the form 
(\ref{slag}), which contains two arbitrary functions $g_1$ 
and $g_2$ with respect to $Y_1=X e^{\lambda \phi/M_{\rm pl}}$ 
and $Y_2=Z e^{\lambda \phi/(2M_{\rm pl})}$. 
In Sec.~\ref{backsec}, we identified the scaling fixed point
(a) as well as the other point (b) relevant to late-time cosmic 
acceleration, without fixing concrete forms of $g_1$ and $g_2$.
We also found that, for models given by the functions 
(\ref{g1e}) and (\ref{g2e}), there exists the scaling 
$\phi$MDE satisfying the relation 
$w_{\rm eff}=\Omega_{\phi}=2Q^2/(3q_s)$. 
Thus, for the existence of $\phi$MDE, we require a nonvanishing 
coupling constant $Q$ associated with the energy transfer.

In Sec.~\ref{backsec}, we studied the background cosmology for 
a concrete model given by the interacting Lagrangian (\ref{Lagcon}).
We showed the existence of $\phi$MDE preceded by 
the radiation fixed point (\ref{pointe}). 
As long as the two conditions (\ref{cond1}) and (\ref{cond2}) 
are satisfied, the $\phi$MDE is followed by the stable fixed 
point (\ref{pointb}) with cosmic acceleration. 
As we observe in Fig.~\ref{fig1}, the field equation of 
state $w_{\phi}$ is close to 1 in the deep matter era 
and it approaches the asymptotic value 
$w_{\phi}=-1+\lambda^2/(3q_s)$ after the 
temporary approach to $-1$ at low redshifts. 
This background dynamics is distinguished from 
the coupled dark energy scenario with the momentum 
transfer alone \cite{Kase:2019mox}.

In Sec.~\ref{persec}, we explored the dynamics of cosmological 
perturbations for the same interacting model studied in Sec.~\ref{backsec}. 
We derived the full linear perturbation equations of motion and 
applied the quasi-static approximation to the modes deep 
inside the sound horizon. 
Under this approximation scheme, the effective gravitational couplings 
for CDM and baryon density contrasts are given by 
Eqs.~(\ref{GCDM}) and (\ref{Gba}), respectively.
We showed that, depending on model parameters,  
both $G_{cc}$ and $G_{cb}$ can be smaller than 
$G$ at low redshifts, while satisfying conditions for the 
absence of ghosts and Laplacian instabilities. 
The weak gravitational interaction of CDM leads to 
the suppression of growth rate 
of total matter density contrast $\delta_m$.
As we observe in Fig.~\ref{fig2}, this property is attributed 
to the momentum transfer arising from the 
coupling $\beta \left( 2X \right)^{1-m/2} Z^m$.

We thus showed that the coupled dark energy and dark matter 
scenario with both energy and momentum exchanges offers an 
interesting possibility for realizing the $\phi$MDE as well as 
the weak gravitational interaction at low redshifts.
The next step is to investigate whether the concrete interacting 
model proposed in this paper can alleviate the problems 
of $H_0$ and $\sigma_8$ tensions present in the $\Lambda$CDM 
model. 
 Since the Lagrangian (5.3) contains additional 
model parameters with respect to those in  $\Lambda$CDM, 
one might expect that our model is hardly better than the
$\Lambda$CDM from the Bayesian statistical point of view.
However, it is known that there are dynamical dark energy models 
that are comparable with  $\Lambda$CDM from the point of view of Bayesian statistics, 
even with more than three  additional free parameters  \cite{Nakamura:2018oyy}.
The detailed observational constraint 
on our interacting dark energy model 
is left for a future publication.

\section*{Acknowledgements}

We thank Antonio De Felice and Ryotaro Kase for useful discussions.
ST is supported by the Grant-in-Aid for Scientific Research Fund of the JSPS No.\,19K03854 and 
MEXT KAKENHI Grant-in-Aid for Scientific Research on Innovative Areas
``Cosmic Acceleration'' (No.\,15H05890).

\section*{Appendix: Phase space for a simplified case}

\begin{figure}[t]
\begin{center}
\includegraphics[width=3.5in]{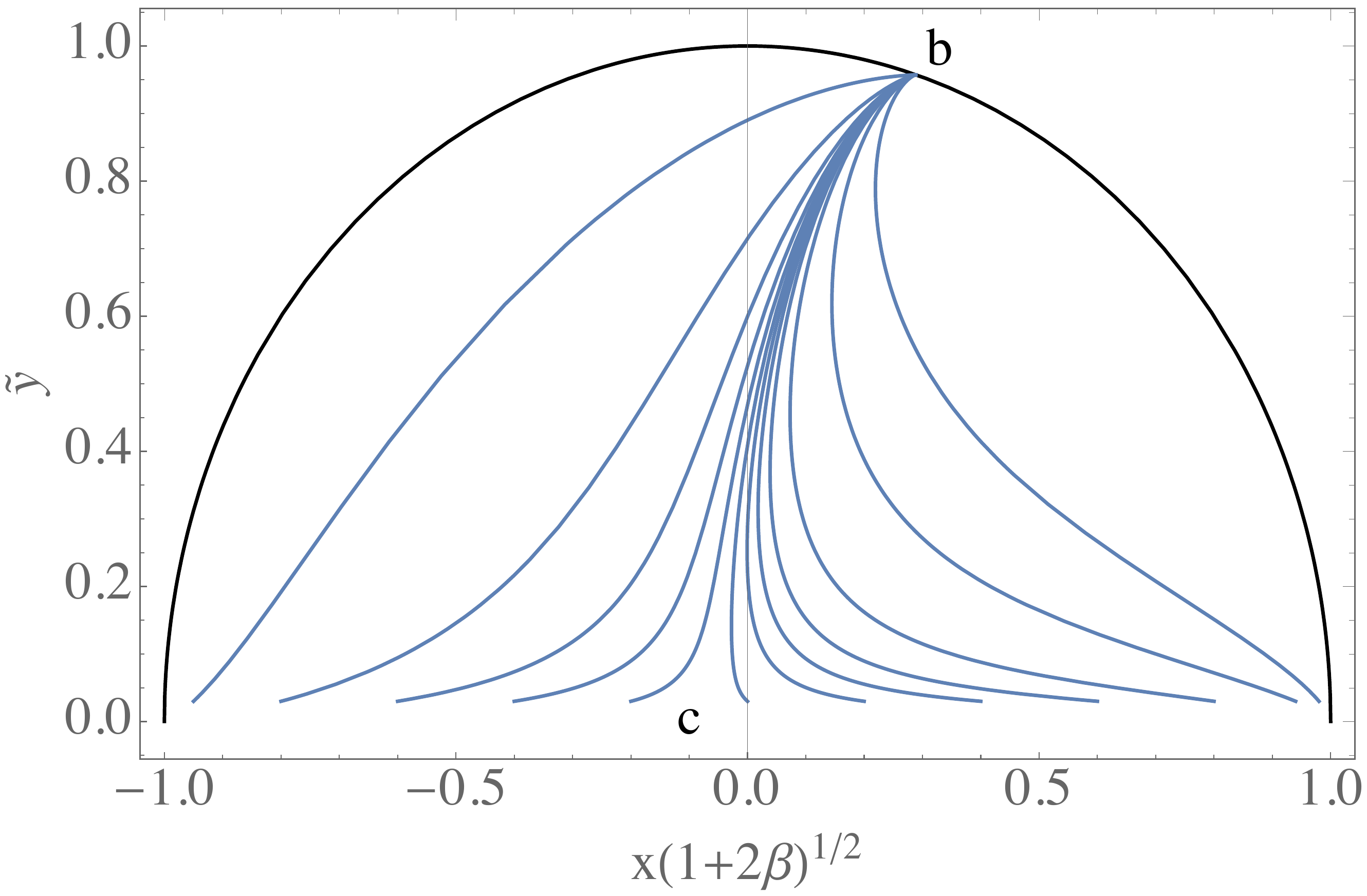}
\end{center}
\caption{ 
Phase-space analysis for the same model parameters as those used 
in Fig.~\ref{fig1}, but with neither baryons nor radiation 
($\Omega_{b}=0$ and $\Omega_{r}=0$). 
The $\phi$MDE saddle point and the final accelerated attractor 
are denoted as c and b, respectively.
\label{fig3}}
\end{figure}

In this Appendix, we express the dynamical Eqs.~(\ref{dxeq})-(\ref{dyeq}) 
in terms of the dimensionless variables $x$, $y$, and $\Omega_m$ 
for the scaling Lagrangian (\ref{slag}) containing 
the functions (\ref{sca}) and (\ref{sca2}).
In doing so, we use the unit $M_{\rm pl}=1$ and notations 
$g_{i,j}=\partial g_i/\partial Y_j$ and 
$g_{i,jk}=\partial^{2}g_{i}/\partial Y_{j}\partial Y_{k}$, 
where $i=1,2$.
Then, the autonomous equations are given by 
\ba
x' &=& 
\frac{\sqrt{6}g_{1}\left(x^{2}\Gamma_{2}+\sqrt{6}g_{2}xy^{4}
+Qy^{4}\Omega_{m}\right)-\sqrt{6}\Omega_{m}\Gamma_{1}}{2\Omega_{m}
\left(\bar{\Gamma}_{1}
+y^{2}g_{1,1}\right)-2g_{1}\left(x^{2}\bar{\Gamma}_{2}
+5x^{2}y^{2}g_{2,1}
+2\sqrt{2}x y^{3}g_{2,2}+g_{2}y^{4} \right)}
+\frac{3x\Theta}{2g_{1}y^{2}}\,,\label{autoge1} \\
y' &=& 
\frac{3\Theta-\sqrt{6}g_{1}\lambda xy^{2}}{2g_{1}y}\,,
\label{autoge2} 
\ea
where 
\ba
\Gamma_{i} &=& \lambda x^2 (\bar\Gamma_i+y^{2}g_{i,1} )
+A_iy^{2}(yg_{i,2}+\sqrt{2} xg_{i,1}) \,,\\
\bar{\Gamma}_{i} &=& 
2x^{2}g_{i,11}+2\sqrt{2}xyg_{i,12}+y^{2}g_{i,22}\,,\\
A_1 &=& \sqrt{2}Qx-\sqrt{3}w_{m}\,,\\
A_2 &=& \sqrt{2}\lambda x+\sqrt{3}\,,\\
\Theta &=& g_{1}\left[ \sqrt{2}x^{3}yg_{2,2}+2g_{2}x^{2}y^2
+y^2\left(w_{m}+1\right)\Omega_{m}+2x^{4}g_{2,1}\right]
-x\,\Omega_{m}\left(2xg_{1,1}+\sqrt{2}yg_{1,2}\right)\,, 
\ea
and
\be
\Omega_{m} =\frac{g_{1}\left(2x^{4}g_{2,1}
+\sqrt{2}x^{3}yg_{2,2}+g_{2}x^{2}y^{2}-y^{2}\right)}
{x\left(2xg_{1,1}+\sqrt{2}yg_{1,2}\right)
-g_{1}y^{2}}\,.
\ee
Let us consider the model given by the functions 
(\ref{eq:simpg1}) and (\ref{eq:simpg2}).
For the perfect fluid, we only take into account CDM with $w_m=0$. 
Then, the autonomous Eqs.~(\ref{autoge1}) and (\ref{autoge2}) 
reduce to
\ba
x' &=& \frac{3}{2} x \left( 2 q_s x^2-2 +\Omega_m \right)
+\frac{\sqrt{6}}{2 q_s} \left( \lambda \tilde{y}^2
-Q \Omega_m \right)\,,\label{autom1}\\
\tilde{y}' &=& \frac{1}{2} \tilde{y} \left( 6 q_s x^2 
-\sqrt{6} \lambda x+3\Omega_m \right)\,,\label{autom2}
\ea
where $\tilde{y}$ is defined by Eq.~(\ref{tildey}), 
and $\Omega_m=1-q_s x^2-\tilde{y}^2$. 
The dynamical system (\ref{autom1})-(\ref{autom2}) 
contains the scaling solution (a) 
characterized by fixed point $(x_c,y_c)=(\sqrt{6}/[2(Q+\lambda)], 
\sqrt{[2Q(Q+\lambda)+3q_s]/[2(Q+\lambda)^2]})$. 
Besides this, there are also the fixed point 
(b) and the $\phi$MDE (c) whose values of $x_c$, $y_c$, 
$\Omega_{\phi}$, $w_{\rm eff}$ are the same 
as those given in Eqs.~(\ref{pointb}) and (\ref{pointc}), 
respectively. We also have the kinetic points (d1) and (d2), 
but they are relevant to neither matter-dominated nor accelerated 
epochs. As long as the two conditions (\ref{cond1}) and 
(\ref{cond2}) are satisfied, the $\phi$MDE is followed 
by point (b) with cosmic acceleration 
[instead of the scaling solution (a)].
In Fig.~\ref{fig3}, we can confirm that, 
for the model parameters used in Fig.~\ref{fig1}, the solutions 
converge to the attractor point (b).

\end{document}